\documentclass[aps,pra,twocolumn,amsmath,amssymb,nofootinbib,superscriptaddress]{revtex4-2}
\usepackage[colorlinks,
           linkcolor=blue,
            anchorcolor=blue,
            allcolors = blue,
            citecolor=blue
            ]{hyperref}

\usepackage{bm}
\usepackage{graphicx}
\usepackage{enumerate}
\usepackage{enumitem}
\usepackage{makecell}
\usepackage{xcolor}

\begin{document}


\title{Simultaneous determination of multiple low-lying energy levels on a superconducting quantum processor}
\date{\today}
\author{Huili Zhang}
\affiliation{Beijing Key Laboratory of Fault-Tolerant Quantum Computing, Beijing Academy of Quantum Information Sciences, Beijing
100193, China}

\author{Yibin Guo}    
\affiliation{Beijing Key Laboratory of Fault-Tolerant Quantum Computing, Beijing Academy of Quantum Information Sciences, Beijing
100193, China}
\affiliation{Institute of Physics, Chinese Academy of Sciences, Beijing 100190, China}
\affiliation{University of Chinese Academy of Sciences, Beijing 101408, China}

\author{Guanglei Xu}
\affiliation{Institute of Physics, Chinese Academy of Sciences, Beijing 100190, China}

\author{Yulong Feng}
\affiliation{Beijing Key Laboratory of Fault-Tolerant Quantum Computing, Beijing Academy of Quantum Information Sciences, Beijing
100193, China}

\author{Jingning Zhang}
   \email{zhangjn@baqis.ac.cn}
\affiliation{Beijing Key Laboratory of Fault-Tolerant Quantum Computing, Beijing Academy of Quantum Information Sciences, Beijing
100193, China}

\author{Hai-feng Yu}
    \email{hfyu@baqis.ac.cn}
\affiliation{Beijing Key Laboratory of Fault-Tolerant Quantum Computing, Beijing Academy of Quantum Information Sciences, Beijing
100193, China}
\affiliation{Hefei National Laboratory, Hefei 230088, China}

\author{S. P. Zhao}
\affiliation{Beijing Key Laboratory of Fault-Tolerant Quantum Computing, Beijing Academy of Quantum Information Sciences, Beijing
100193, China}
\affiliation{Institute of Physics, Chinese Academy of Sciences, Beijing 100190, China}


\begin{abstract}

Determining the ground and low-lying excited states is critical in numerous scenarios. Recent work has proposed the ancilla-entangled variational quantum eigensolver (AEVQE) that utilizes entanglement between ancilla and physical qubits to simultaneously tagert multiple low-lying energy levels. In this work, we report the experimental implementation of the AEVQE on a superconducting quantum cloud platform, demonstrating the full procedure of solving the low-lying energy levels of the H$_2$ molecule and the transverse-field Ising models (TFIMs). We obtain the potential energy curves of H$_2$ and show an indication of the ferromagnetic to paramagnetic phase transition in the TFIMs from the average absolute magnetization. Moreover, we investigate multiple factors that affect the algorithmic performance and provide a comparison with ancilla-free VQE algorithms. Our work demonstrates the experimental feasibility of the AEVQE algorithm and offers a guidance for the VQE approach in solving realistic problems on publicly-accessible quantum platforms.
\end{abstract}

\maketitle

\section{Introduction}

\begin{figure*}
    \centering
    \includegraphics[width=0.99\textwidth]{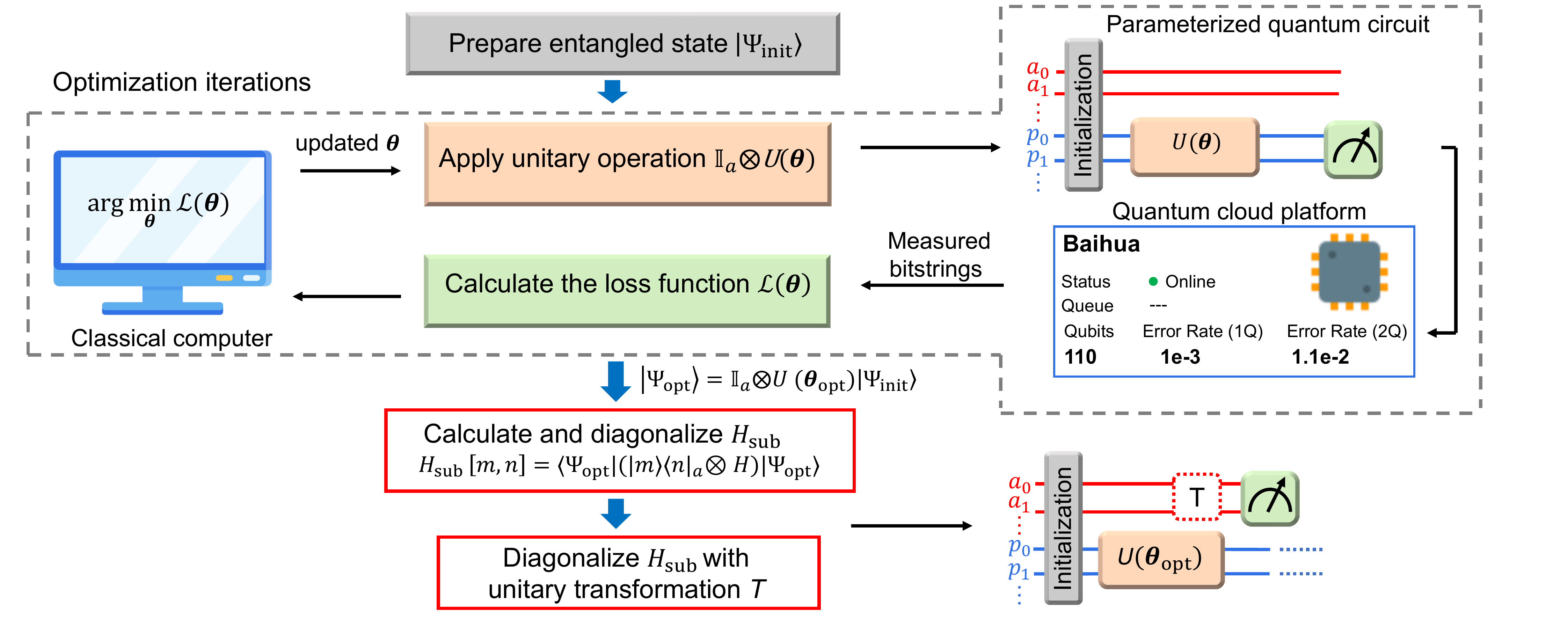}
    \caption{Schematic diagram of AEVQE. The quantum circuit consists of ancilla qubits $a_i$ and physical qubits $p_i$. First, the ancilla and physical qubits are initialized to entangled states. In the optimization iterations, the variational circuits with parameters $\bm{\theta}$ are executed on the quantum processor. The measured results of the physical qubits (noted as bitstrings)  are fed into the classical computer for searching $\bm{\theta}_{\text{opt}} = \text{arg}\mathop{\text{min}}\limits_{\bm{\theta}}\mathcal{L}(\bm{\theta})$. After optimization, $\bm{\theta}_{\text{opt}}$ is obtained and applied to calculate $H_{\text{sub}}$. Finally, unitary transformation $T$ is applied to diagonalize $H_{\text{sub}}$ and determine the eigenenergies and eigenstates.}   
    \label{Fig1}
\end{figure*}

Quantum computers have potential computational advantages over classical computers~\cite{ lloyd1996universal, georgescu2014quantum, arute2019quantum, zhong2020quantum, madsen2022quantum} in various applications such as integer factoring \cite{shor1999polynomial}, quantum simulation \cite{aspuru2005simulated, yuan2020quantum, mcardle2020quantum, bauer2023quantum, gerritsma2010quantum, bloch2012quantum}, and quantum artificial intelligence \cite{jordan2015machine, carleo2019machine, deng2017machine, bausch2024learning, alexeev2024artificial, koutromanos2024control}.  Variational quantum algorithms (VQAs) are one of the leading algorithms for achieving quantum advantage in current noisy intermediate-scale quantum (NISQ) devices~\cite{cerezo2021variational,bharti2022noisy,kandala2017hardware, sweke2020stochastic, tang2021qubit}. The first VQA, the variational quantum eigensolver (VQE), is developed to find the ground state of a specific Hamiltonian $H$~\cite{google2020hartree, peruzzo2014variational, liu2019variational, nam2020ground, hempel2018quantum,wang2023electronic}, which utilizes a parameterized ansatz quantum circuit $U(\bm{\theta})$ to generate an ansatz state $|{\psi(\bm\theta)}\rangle$. By minimizing the expectation value $\langle{\psi(\bm\theta)}|H|{\psi(\bm\theta)}\rangle$ (i.e., the loss function) via iterative optimization of the circuit parameters $\bm\theta$, the ground state of the target Hamiltonian can be found. The parameterized  quantum circuit in the algorithm enables the reduction of the quantum circuit depth, which is the key to realizing quantum advantage for NISQ devices.

In addition to ground state eigenenergies, the excited states also play critical roles in various studies like the prediction and analysis of chemical reactions and phase transitions. Multiple VQE approaches have been extended to calculate the low-energy excited states, including the variational quantum deflection (VQD), the non-weighted and weighted subspace search VQE (SSVQE), and the multistate contracted VQE (MCVQE)\cite{ibe2022calculating, higgott2019variational, jones2019variational, benavides2024quantum, smart2024many, jouzdani2021method, nakanishi2019subspace, han2024multilevel, yalouz2021state, parrish2019quantum,dutta2025qumode,guo2024concurrent}. In the VQD, the $K$ eigenvalues $E_{0},...,E_{K-1} $ are computed recursively from lowest to highest, with the loss function for finding the $k$th energy level given by $\mathcal{L}_{k}(\bm\theta) = \langle\psi_{k}|H|\psi_{k}\rangle + \sum_{i=0}^{k-1}\beta_{i}|\langle\psi_k|\psi_i\rangle|^{2}$, where $\beta_{i}$ are chosen to be sufficiently large to ensure the orthogonality of the eigenstates~\cite{ibe2022calculating, higgott2019variational, jones2019variational, benavides2024quantum, smart2024many, jouzdani2021method}. In the weighted SSVQE, $K$ orthogonal initial states $|\psi _{i}\rangle$ are prepared and evolved to $U(\bm{\theta})|\psi_{i}\rangle$ through the parameterized circuit. The loss function is expressed by the weighted sum of the $K$ orthogonal output states in the form $\mathcal{L}(\bm{\theta}) = \sum_{i=0}^{K-1}w_{i}\langle\psi_{i}|U^{\dagger}(\bm{\theta})HU(\bm\theta)|\psi_{i}\rangle$\cite{nakanishi2019subspace, han2024multilevel, yalouz2021state}. Both non-weighted SSVQE and MCVQE employ uniform weighting schemes; the former targets the $k$th excited state, whereas the latter extracts the  $K$ lowest-lying eigenvalues. However, the realization of these algorithms on quantum hardware is still lacking.

Recently, a VQE approach capable of simultaneously determining multiple eigenstates has been proposed \cite{xu2023concurrent}. The method uses a set of ancilla qubits to construct maximally entangled states with corresponding qubits in the system, so that multiple final eigenstates evolved from the initial orthogonal states can be obtained from different ancillary states. We refer to this approach as ancilla-entangled VQE (AEVQE). In this work, we present the experimental demonstration of AEVQE with the superconducting quantum processor \textit{Baihua} on the platform \textit{Quafu SQC}~\cite{baqisQuafuSuperconducting}. We apply the algorithm to simulate the $\text{H}_{2}$ molecule and transverse field Ising models (TFIMs). For the $\text{H}_{2}$ molecule, we obtain the H-H bond distance dependence of two eigenenergies, with an average energy difference of 0.027 Hartree for the first excited state. For the TFIMs, combined with symmetry verification methods, we calculate four eigenenergies for the three-spin system and two eigenenergies for the five-spin system, with average differences of 0.029 and 0.099 for the high-lying excited states, respectively. Then, we show indications of the phase transition of the model by calculating the average absolute magnetization. Moreover, we carry out simulations on TFIMs to analyze the factors that affect the optimization efficiency, including the system size, the choice of classical optimizer, and hyperparameters. Finally, we compare our algorithm with the weighted SSVQE and MCVQE algorithm, and especially, analyze their shot budget in the optimization stage. Our results showcase the capability and challenges of AEVQE for determining multiple eigenenergies and eigenstates of many-body systems on quantum hardware. 

\begin{figure}[t]
	\centering
	\includegraphics[width=0.5\textwidth]{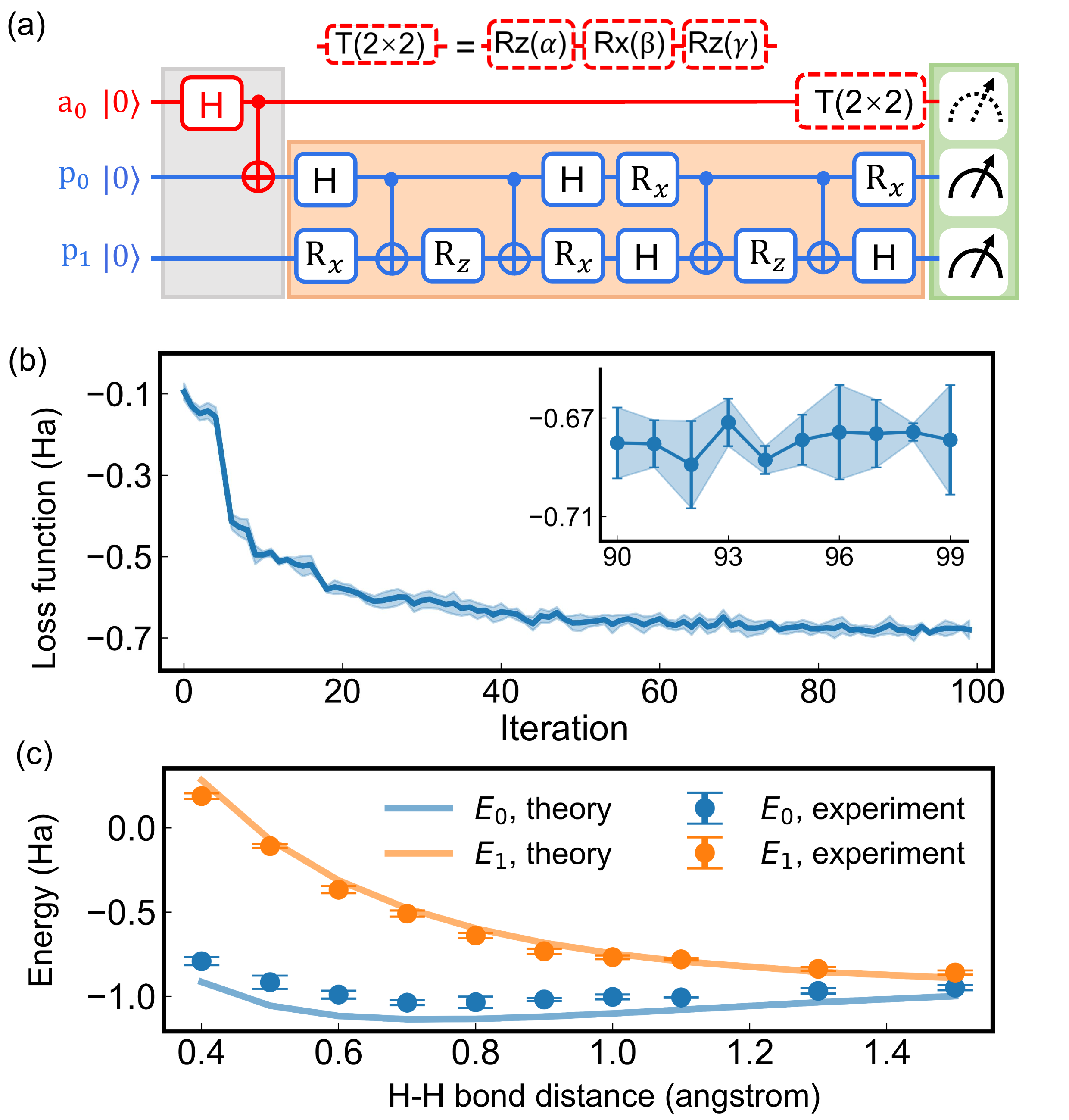}
	\caption{Calculation of the ground and first excited state energies $E_0$ and $E_1$ of the $\text{H}_{2}$ molecule. (a) The circuit schematic. The gray and orange areas represent the initialization and variational circuits, respectively. $H$ represents the Hadamard gate. $R_x$ and $R_z$ are the rotations along the $x$ and $z$ axes of the Bloch sphere, with the rotation angles to be optimized. The unitary transformation $T$ is applied to diagonalize $H_\text{sub}$ after the loss function convergence. (b) The optimization of the loss function at H-H bond distance $d = 0.6$ angstrom. The inset shows the result of the final ten iterations. (c) The experimental energy potentials $E_0$ (blue dots) and $E_1$ (orange dots) as a function of the H-H bond distance. The lines are the corresponding theoretical results. The error bars represent the standard error of the average energy. All energies are in the unit of Hartree.}
	\label{Fig2}
\end{figure}  

\section{Scheme} 

To determine $K$ eigenenergies and eigenstates of $H$ (embedded in $N_{p}$ physical qubits), additional $N_{a}$ ancilla qubits $( N_{a} < N_{p})$ are required. To maximize the utilization of ancilla qubits, we always set $K = 2^{N_a}$. The process of AEVQE is illustrated in Fig. \ref{Fig1} and is described as follows: 
\begin{enumerate}[label= (\arabic*),itemsep=0pt]
\setlength{\itemindent}{1.1em}
\setlength{\itemsep}{0pt}
\item Choose $N_{a}$ physical qubits and prepare a maximally entangled state between the $i$th ancilla qubit and $i$th physical qubit to construct an initial state $|\Psi_{\rm{init}}\rangle = \frac{1}{\sqrt{K}}\sum_{m=0}^{K-1}|m\rangle_a|\psi_m\rangle_p$, with $\{|m\rangle \}_{m=0}^{K-1} $being the computational basis and $\langle m|n \rangle = \delta_{m,n} $. The subscript $a(p)$ denote the ancilla (physical) qubit, respectively.

\item Apply the unitary operation $\mathbb{I}_{a} \mathop{\otimes}U(\bm{\theta})$ to the initialized system, where $\mathbb{I}_{a}$ represents the identity operator on the ancilla qubits, and $U(\bm{\theta})$ represents the parametrized circuit acting on the physical qubits, with $\bm{\theta}$ being the variational parameter. 

\item Measure the physical qubits and evaluate the loss function
\begin{eqnarray}
    \mathcal{L(\bm{\theta})} &= &\frac{1}{K}\sum_{m=0}^{K-1}\ \langle\psi_m|_pU^{\dagger}(\bm\theta)HU(\bm{\theta})|\psi_m\rangle_p
\end{eqnarray}

which sets an upper bound on the average of the $K$ low-lying eigenenergies of  $H$.  The gradient of the loss function is fed to the classical optimizer to obtain the updated $\bm{\theta}$. Repeat Steps (1-3) until the loss function converges.

\item Measure the ancilla qubits and the physical qubits to calculate $H_\text{sub}$, with the matrix element given by 
\begin{equation}
    H_{\text{sub}}[m,n] = \langle\Psi_{\text{opt}}|(|m\rangle\langle n|_{a}\otimes H)|\Psi_{\text{opt}} \rangle,
\end{equation}
and $|\Psi_{\text{opt}}\rangle = \mathbb{I}_{a}\otimes U(\bm{\theta}_{\text{opt}})|\Psi_{\text{init}}\rangle$ being the optimal final state.
\item Diagonalize $H_{\text{sub}}$ such that $H_{\text{sub}} = T^{\dagger}DT$ with $D = \text{diag}(E_0,...E_{K-1})$  and $T$ being the unitary transformation that maps the eigenbasis of $H_{\text{sub}}$ to the computational basis. The eigenstate $|E_m\rangle$ can be prepared on the physical qubits by projectively measuring the ancilla qubits and post-selecting the outcome $m$.
\end{enumerate}
 
In this experiment, we choose the simultaneous perturbation stochastic approximation (SPSA)~\cite{spall1992multivariate}, a gradient-based  classical optimizer, to minimize the loss function in Eq. (1). At the $k$th iteration step, the stochastic approximation of the gradient of the loss function
\begin{equation}
    \bm{g}_{k}(\bm{\theta}_{k}) = \bigg(\frac{\mathcal{L}_k^{+} - \mathcal{L}_{k}^{-}}{2\epsilon\delta_{k0}}, ...,  \frac{\mathcal{L}_k^{+} - \mathcal{L}_{k}^{-}}{2\epsilon\delta_{kp}}\bigg)^{T}
\end{equation}
is obtained by evaluating the loss function $\mathcal{L}_k^{\pm} = \mathcal{L} (\bm{\theta}_{k}\pm\epsilon\bm{\Delta}_k)$ twice, where $\epsilon$ quantifies the pertubation strength and $\bm{\Delta}_k = (\Delta_{k0},....,\Delta_{kp})^{T}$ is a random perturbation vector with each $\Delta_{ki}$ independently chosen from $\pm1$ with equal probability. Then the  variational parameters are updated using $\bm{\theta}_{k+1} = \bm{\theta}_{k} + \eta \bm{g}_{k}(\bm{\theta}_{k})$, where $\eta$ is the learning rate.  In this experiment, we set the perturbation $\epsilon = 0.1$ and the learning rate $\eta = 0.2$.  

Our experiment is carried out with the quantum processor \textit{Baihua} on the \textit{Quafu SQC} platform (see Fig.~\ref{Fig1}). For the measurement errors, we apply readout error mitigation to the classical measurement outcomes. To reduce the sampling errors, the measurements are repeated by 15 $\times$ 1024 times in each iteration. In cases where the TFIMs are simulated, we apply symmetry verification to reduce gate errors.  These techniques are essential for the stabillity and accuracy of the quantum-cloud-based AEVQE implementation. 

\section{Simulation of the $\mathbf{H_2}$ molecule}

To start, we use AEVQE to experimentally calculate the ground and first excited state energies of the $H_2$ molecule. Two physical qubits are required to implement the variational circuit of the unitary coupled-cluster generalized singles and doubles (UCCGSD) ansatz \cite{hong2024Refining}. The Hamiltonian $H$ \cite{mcclean2020openfermion,sun2018pyscf} is given by 
\begin{equation}
H = c_0 + c_1 Z_0 + c_2 X_0 + c_3 Z_0Z_1 + c_4 X_0X_1,
\end{equation}
with the coefficients $c_j$ $(j=1,2,3,4)$ depending on the H-H bond distance $d$, and $X_i$, $Y_i$, and $Z_i$ being the Pauli operators for the $i$th qubit. Introducing one ancilla qubit enables the simultaneous determination of the two lowest eigenenergies $E_0$ and $E_1$. The quantum circuit is shown in Fig. \ref{Fig2}(a). In the initialization step, the ancilla qubit and the first physical qubit are prepared in the Bell state $(|00\rangle+|11\rangle)/\sqrt{2}$, and the second physical qubit is prepared in $|0\rangle$. In the variational step, the unitary circuit is applied to two physical qubits, and the rotation angles of $R_x$ and $R_z$ are variational parameters. In the measurement step, the physical qubits are measured to calculate the loss function. The loss function minimization procedure at the H-H bond distance $d=0.6$  angstrom is shown in Fig. \ref{Fig2}(b), which show good convergence within 100 iteration steps.  

After optimizing, the initial orthogonal states are transformed to the subspaces spanned by the low-lying eigenstates, and the corresponding eigenenergies can be obtained by diagnonalizing the subspace Hamiltonian $H_\text{sub}$.  The matrix elements of $H_{\text{sub}}$ are measured as the expectation values of the observables listed in Table \ref{table1}. By varying the H-H bond distance, we experimentally obtain the potential energy curves $E_0$ and $E_1$, which are shown in Fig. \ref{Fig2}(c). The average deviations between experiment and theory are 0.098 Hartree for $E_{0}$ and 0.027 Hatree for $E_1$. The results demonstrate that the computational accuracy of the excited state eigenenergy is not affected by the errors in the ground state eigenenergy.  

\begin{table}[t]
    \centering
    \caption{The observables for  the matrix elements of $H_{\text{sub}}$ for calculating the ground and first excited state energies of the $\text{H}_2$ molecule. $X$,$Y$,$Z$ are Pauli operators, and $I$ represents the identity operator.}
    \setlength{\tabcolsep}{26pt}
    \begin{tabular}{c|c}
      \toprule
      Observable   &  Matrix element\\\hline
      $(I+Z) \otimes H$    & $H_{\text{sub}}[0,0]$ \\
      $(X+iY) \otimes H$    & $H_{\text{sub}}[0,1]$ \\
      $(X-iY) \otimes H$    & $H_{\text{sub}}[1,0]$ \\
      $(I-Z) \otimes H$    & $H_{\text{sub}}[1,1]$ \\\hline
    \toprule
    \end{tabular}
    \label{table1}
\end{table}

\begin{figure*}[t]
	\centering
	\includegraphics[width=0.95\textwidth]{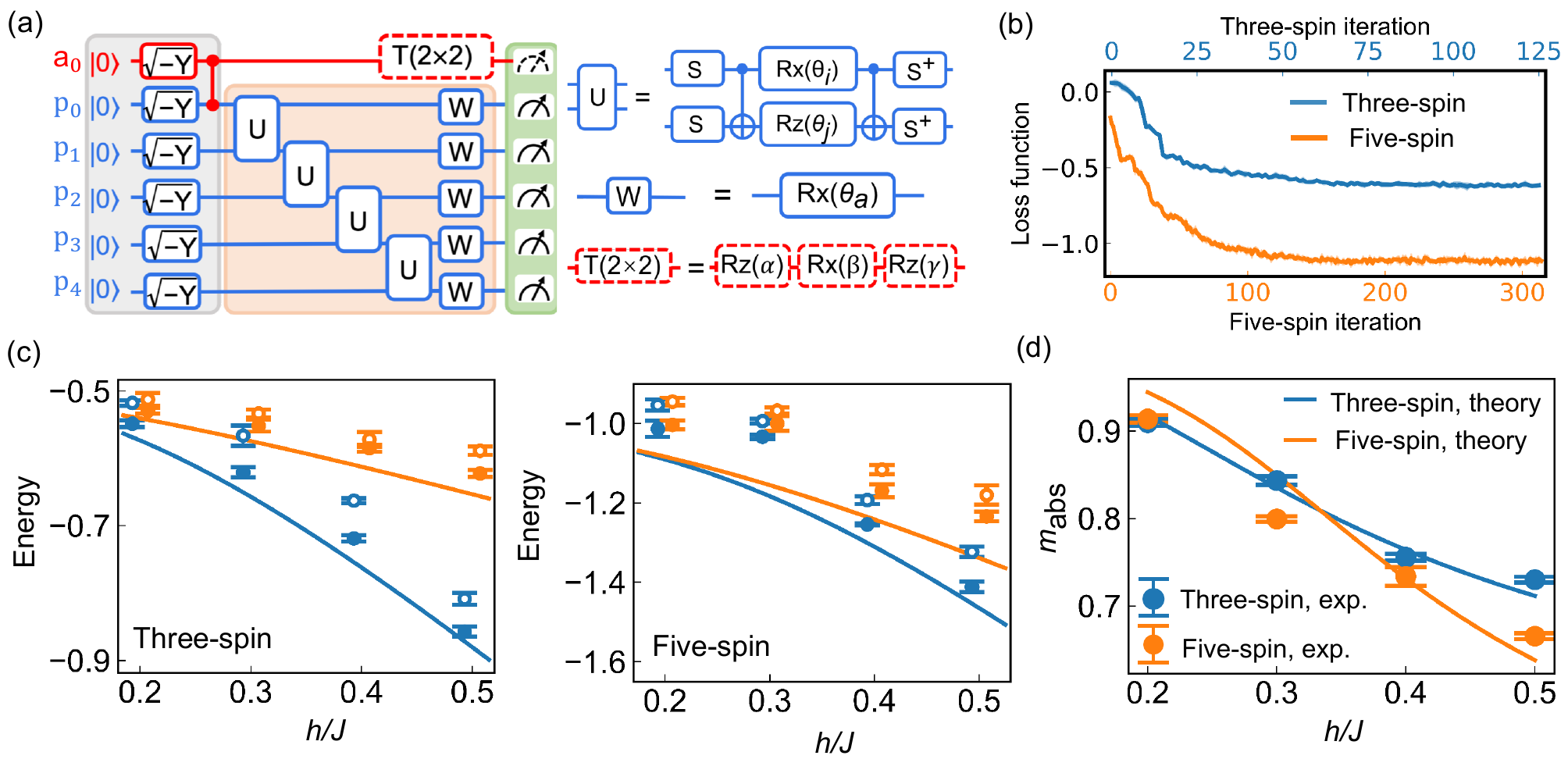}
	\caption{Calculation of the ground and first excited state energies $E_0$ and $E_1$ of the transverse field Ising models with one ancilla qubit. (a) The circuit schematic for the five-spin system.  Decompositions of $U$, $W$, and $T$ are shown on the right. $S = R_z(\pi/2)$ is the phase gate and $\sqrt{-Y} = R_y(-\pi/2)$. (b) The optimization of the loss function for $h/J = 0.4$. The blue and orange lines represent the results of the  three- and five-spin TFIMs, respectively. (c) The ground and first excited state energies ($E_0$ , $E_1$) of the three- and five-spin TFIMs obtained in experiment. The open circles and solid circles represent the raw energies and the symmetry-verified energies,respectively. The error bars represent the standard error of the average energy. (d) The experimental (dots) and theoretical (lines) average absolute magnetization $m_{\text{abs}}$ of the three- and five-spin TFIMs.}
	\label{Fig3}
\end{figure*}

\section{Transverse field Ising Model}

TFIMs are widely studied in diverse fields of many-body quantum physics~\cite{
pfeuty1970one, heyl2013dynamical, mondaini2016eigenstate, schmitt2022quantum, li2023probing}. The Hamiltonian of the one-dimensional $N_p$-spin TFIM is described by:
\begin{equation}
    H = -J/2\sum_{i=0}^{N_p-2} Z_{i}Z_{i+1}+h\sum_{i=0}^{N_p-1}X_i,  \label{TFIM}
\end{equation}
where $X_i$ and $Z_i$ are the Pauli operators acting on the $i$th qubit, $J$ represents the strength of the spin-spin interaction, and  $h$ is the strength of the external field. 

We use one ancilla qubit to calculate the ground-state and the first-excited-state energies of the TFIMs with three and five spin sites. The circuit of AEVQE with one ancilla qubit and five physical qubits is shown in  Fig. \ref{Fig3}(a) on the left side, with the decompositions of the matrices $U$, $W$, and $T$ shown on the right. In the calculation, the ancilla qubits and the physical qubits are initialized in the $(|0\rangle-|1\rangle)/\sqrt{2}$ state. The control-Z (CZ) gate is applied to create entanglement between each ancilla qubit and its adjacent physical qubit. The optimization results of the loss function with $h/J=0.4$ for the three- and five-spin TFIMs are shown in Fig.~\ref{Fig3}(b).

After the convergence of the loss function, we use the optimized parameters to calculate the eigenenergies. It is noteworthy that the Hamiltonian $H$ in Eq.~(\ref{TFIM}) commutes with the parity operator $P_X = \bigotimes_{i =0}^{N_p-1}X_{i}$, while the ground state $|E_0\rangle$ and the first excited state $|E_1\rangle $ are the eigenstates of $P_X$ with eigenvalues of +1 and -1, respectively. Hence, symmetry verification can be applied to reduce errors \cite{cai2023quantum,bonet2018low}.  The symmetry-verified eigenenergies $E_{0,1}$ and density matrices $\rho^{\text{sym}}_{0,1}$ are given by
\begin{eqnarray}
E_{0,1} = \frac{\text{Tr}[HP_{\pm}\rho_{0,1}^{\text{raw} }P_{\pm}]}{\text{Tr}[P_{\pm}\rho_{0,1}^{\text{raw}} P_{\pm}]},\ \     \rho_{0,1}^{\text{sym}} = \frac{P_{\pm}\rho_{0,1}^{\text{raw} } P_{\pm}}{\text{Tr}[P_{\pm}\rho_{0,1}^{\text{raw} } P_{\pm}]},
\end{eqnarray} 
where $P_{\pm} = (1\pm P_X)/2$ are the projectors onto the corresponding symmetry subspaces and $\rho^{\text{raw}}_{0,1}$ are the density matrices obtained directly from the measurement outcomes. The acceptance rates $\text{Tr}[P_{\pm}\rho^{\text{raw}}_{0,1} P_{\pm}]$ are listed in Table \ref{table2}. The experimental eigenenergies obtained after symmetry verification are shown in Fig. \ref{Fig3}(c). The average energy difference for the three- and five-spin TFIMs are 0.0205 and 0.0987, respectively, showing that the error increases with system size.

\begin{table}[b]
	\centering
	\caption{The acceptance rates in the symmetry verification.}
	\setlength{\tabcolsep}{8pt}
		\begin{tabular}{c|c|c|c|c}
			\toprule
			$h/J$   &  \thead{Three-spin \\$E_0$}& \thead{Three-spin\\ $E_1$}& \thead{Five-spin \\$E_0$} & \thead{Five-spin \\$E_1$}\\\hline
			0.2 & 0.944 & 0.968& 0.941&0.942\\
			0.3 & 0.912 & 0.968& 0.962&0.968\\
			0.4 & 0.922 & 0.978& 0.951&0.954\\
			0.5 & 0.943 & 0.946& 0.937&0.956\\\hline
			\toprule
		\end{tabular}
		\label{table2}
	\end{table}
	
	\begin{figure*}[t]
		\centering
		\includegraphics[width=0.99\textwidth]{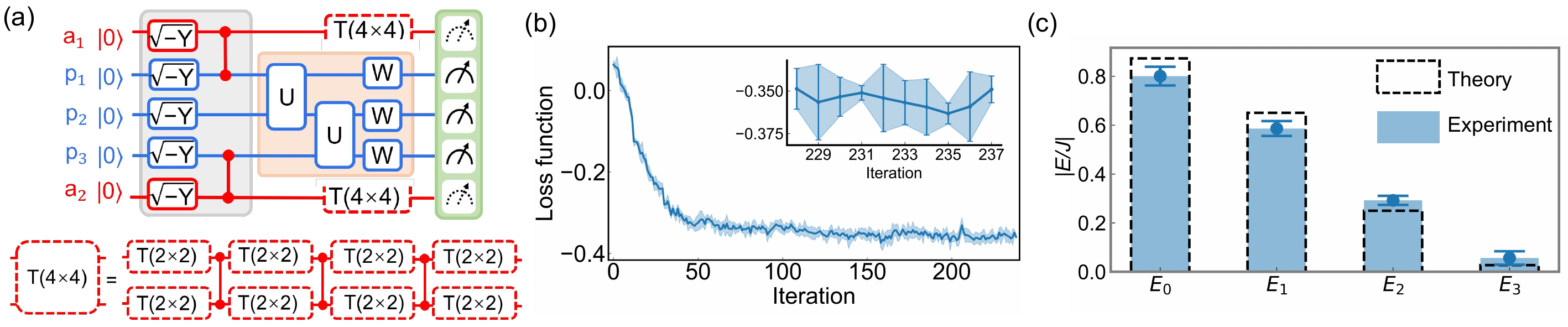}
		\caption{Calculation of the lowest four eigenenergies $E_0$, $E_1$, $E_2$, and $E_3$ of the transverse field Ising model with two ancilla qubits. (a) The circuit schematic for the three-spin system. The transformation matrix $T(4\times4)$ is decomposed into single qubit rotations and CZ gates, as shown at the bottom. The unitary operators $U$, $W$, and $T(2\times2)$ are the same as those in Fig.~\ref{Fig3}(a). (b) The optimization of the loss function for $h/J = 0.5$. (c) The experimental eigenenergies $E_0$, $E_1$, $E_2$, and $E_3$ of the three-spin TFIM with $h/J = 0.5$. The dashed-line rectangles are the corresponding theoretical results. The error bars represent the standard errors of the average energy.}
		\label{Fig4}
	\end{figure*}
	
Beyond eigenenergies, AEVQE can directly prepare the associated eigenstates, thereby facilitating measurement of  their physical properties.  Conventionally, we use the computational basis $|0\rangle$ and $|1\rangle$ of each qubit to represent the up state and down state  of each spin, respectively. The ground-state average absolute magnetization $m_{\text{abs}} $, defined  as 
\begin{equation}
m_{\text{abs}} = \frac{1}{N_p}\sum_{s=0}^{N_p}|N_p-2s|P(s),
\end{equation}
where $P(s)$ is the measured probability of  finding $s$ spins in the  down state.  This quantity serves as the order parameter for the ferromagetic-paramagnetic phase transition \cite{islam2011onset}. In the limit of $h/J \to 0$, all spins are ordered by the Ising interaction, such that the order parameter $m_{\text{abs}}  = 1$. In the opposite limit of $h/J \to \infty$, the system is mostly polarized by the external magnetic field, leading to $m_{\text{abs}} = 0$. In the thermodynamic limit, this order parameter shows a non-analytic behavior at $h/J=0.5$, i.e, the quantum phase transition \cite{binder1981critical,fisher1972scaling,islam2011onset}. For finite systems, the phase transition manifests as a sharpening of $m_{\text{abs}} $ with increasing system size, as shown in Fig. \ref{Fig3}(f).

\begin{table*}[!tp]
	\centering
	\caption{The observables for  the matrix elements of $H_{\text{sub}}$ for calculating four eigenenergies of the TFIM. $X$, $Y$, $Z$ are Pauli operators, and $I$ represents the identity operator.}
	\setlength{\tabcolsep}{20pt}
		\begin{tabular}{c|c||c|c}
			\toprule
			Observable   &  Matrix element &  Observable   &  Matrix element\\\hline
			$(I+Z)\otimes (I+Z)  \otimes H$    & $H_{\text{sub}}[0,0]$ &  $(X-iY)\otimes (I+Z) \otimes H$ & $H_{\text{sub}}[2,0]$ \\
			$(I+Z)\otimes (X+iY) \otimes H$    & $H_{\text{sub}}[0,1]$ & $(X-iY)\otimes (X+iY) \otimes H$ & $H_{\text{sub}}[2,1]$ \\
			$(X+iY)\otimes(I+Z)  \otimes H$    & $H_{\text{sub}}[0,2]$ & $(I-Z)\otimes(I+Z) \otimes H$  & $H_{\text{sub}}[2,2]$\\
			$(X+iY)\otimes(X+iY) \otimes H$    & $H_{\text{sub}}[0,3]$ & $(I-Z)\otimes(X+iY)  \otimes H$  & $H_{\text{sub}}[2,3]$ \\
			$(I+Z)\otimes (X-iY) \otimes H$    & $H_{\text{sub}}[1,0]$ &  $(X-iY)\otimes (X-iY) \otimes H$ & $H_{\text{sub}}[3,0]$ \\
			$(I+Z)\otimes (I-Z)  \otimes H$    & $H_{\text{sub}}[1,1]$ &  $(X-iY)\otimes (I-Z)  \otimes H$  & $H_{\text{sub}}[3,1]$ \\
			$(X+iY)\otimes(X-iY) \otimes H$    & $H_{\text{sub}}[1,2]$ & $(I-Z)\otimes(X-iY) \otimes H$  & $H_{\text{sub}}[3,2]$\\
			$(X+iY) \otimes(I-Z)  \otimes H$   & $H_{\text{sub}}[1,3]$ &  $(I-Z)\otimes(I-Z)  \otimes H$ & $H_{\text{sub}}[3,3]$ \\\hline
			\toprule
		\end{tabular}
		\label{table3}
	\end{table*}
	
We next calculate the lowest four eigenenergies of the three-spin TFIM with $h/J = 0.5$ by introducing two ancilla qubits. As illustrated in Fig. \ref{Fig4}(a), two pairs of ancilla and physical qubits are initialized in entangled states, while the variational circuit retain same structure as in Fig. \ref{Fig3}(a). The convergence of the loss function is shown in Fig. \ref{Fig4}(b). Compared to the  three-spin TFIM with one ancilla qubit, while the number of optimization parameters remains unchanged, the average number of iterations nearly doubles. This indicates that as  the dimension of the target low-lying eigenspace increases, the optimization landscape becomes more complex. After the convergence of the loss function, we measure both  physical and ancilla qubits to reconstrust $H_\text{sub}$, with the following observables listed in Table \ref{table3}. The eigenenergies obtained by diagonalizing $H_\text{sub}$ are shown in Fig. \ref{Fig4}(c). The bias of the experimental eigenenergies $E_0, E_1, E_2 $ and $E_3$ is 0.072,  0.064, -0.042, and -0.029, respectively. Notably, the error magnitude does not significantly  increase for higher energy levels, demonstrating the feasibility of AEVQE in calculating higher excited states. 

\section{Scaling and optimization of AEVQE}
	
To investigate the effects of system size and classical optimizers on AEVQE, we perform numerical simulations of the TFIMs at $h/J=0.5$ using \textit{Qiskit} \cite{qiskit}. The schematic of the quantum circuit consisting of $N_a$ ancilla qubits and $N_p$ physical qubits is shown in  Fig. \ref{Fig5}(a). In the simulations, we introduce depolarization noise with error rates of 0.001 for single-qubit gates and 0.01 for CZ gates. For each case, the optimization stage is repeated by 100 trials, with randomly chosen initial variational parameters. The convergence criterion is defined as the difference of the loss function and the exact average energy $E= (E_0+E_1)/2$ falls below 0.05 in the unit of $J$.

\subsection{Effect of system size}

\begin{figure}[b]
	\centering
	\includegraphics[width=0.5\textwidth]{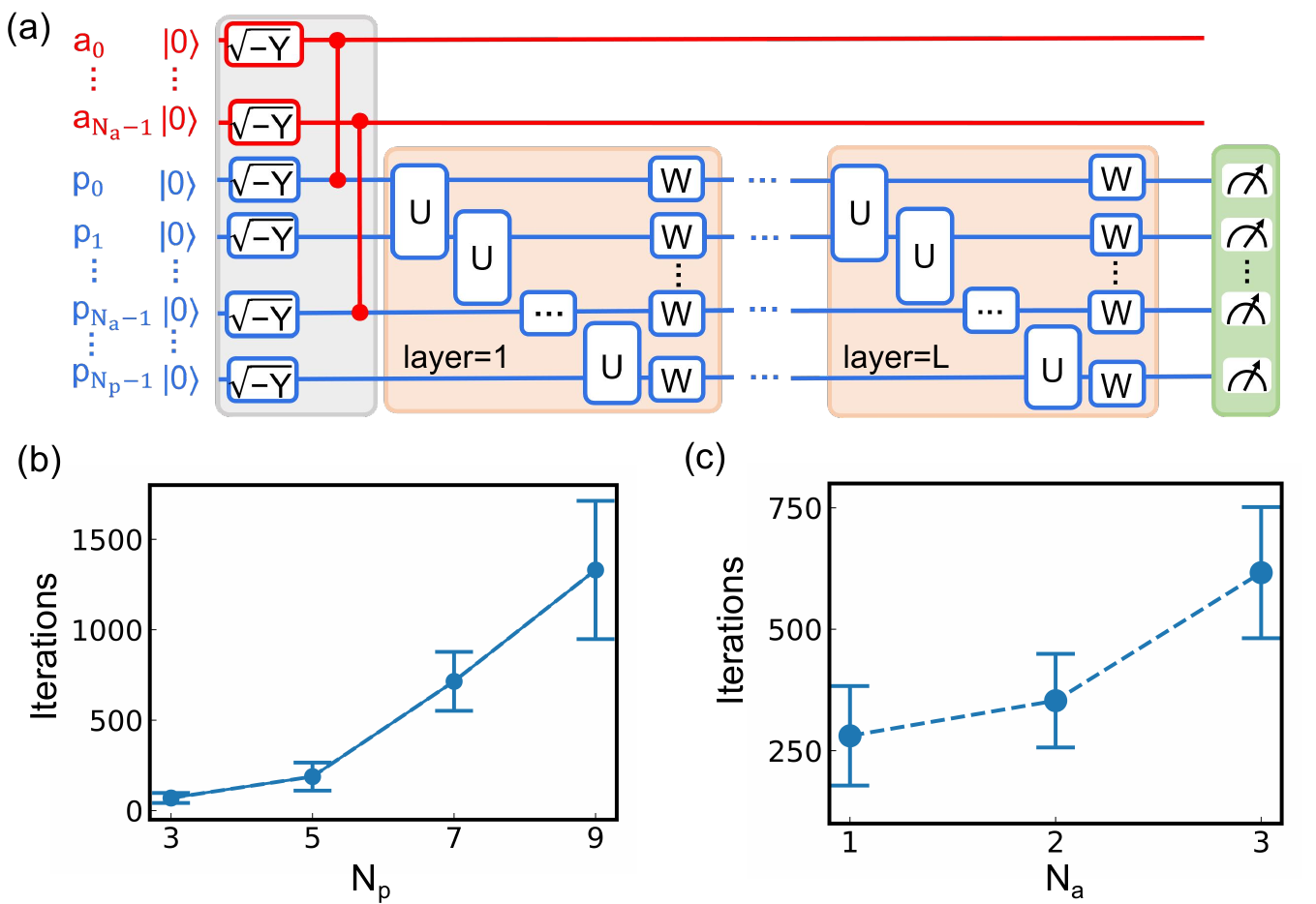}
	\caption{(a) The AEVQE circuit for simulating the $N_p$-spin TFIMs, consisting of $N_p$ physical qubits and $N_a$ ancilla qubits. (b) The number of iterations as a function of the number of physical qubits, with fixed one ancilla qubit. (c) The number of iterations in simulating the five-spin TFIM with different ancilla qubit numbers. The error bars represent the standard errors of the average iterations.}
	\label{Fig5}
\end{figure}

The number of physical qubits directly influences the circuit depth and the number of optimization parameters, hence strongly affecting the optimization stage of AEVQE.  We increase the number of physical qubits from three to nine while keeping one ancilla qubit. Fig. \ref{Fig5}(b) illustrates the average number of iterations as a function of the number of physical qubits. As the system scales up, more iterations are required to achieve the same accuracy.For the three-spin TFIM, the average number of iterations is $70\pm28$, while for the nine-spin TFIM, the average number of iterations is $1330\pm382$, almost 19 times that required for the three-spin case. We also note that for the nine-spin TFIM, the number of successful trials is only 15, approximately one-sixth of the total trials. This reduction of can be explained by the barren plateau phenomenon, in which  the gradient vanishes exponentially with the qubit numbers in the variational ansatz, systematically degrading the trainability of the optimization landscape \cite{stilck2021limitations, sharma2022trainability,wang2021noise,mcclean2018barren,arrasmith2021effect}.

Next, we focus on the five-spin TFIM, and increase the number of ancilla qubits up to three to test the effect on the optimization performance. In the case of $N_a =1,2$, applying only one variational layer yields the average iterations of $280 \pm 102$ and $353 \pm 96$, respectively. For $N_a =3$, obtaining the same accuracy requires two variational layers. Across 22 successful trials, the average number of iterations increases to $616 \pm 134$. This results are summarized in Fig. \ref{Fig5}(c), indicating that as the number of ancilla qubits,greater circuit expressibility is needed, which can be achieved  by increasing circuit depth and thereby expanding the parameter space. 

\subsection{Effect of classical optimizers}

\begin{figure}[b]
	\centering
	\includegraphics[width=0.45\textwidth]{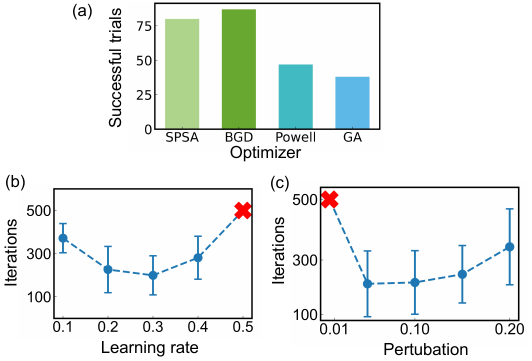}
	\caption{The effect of classical optimizers and hyperparameters on the optimization efficiency of AEVQE. The results are obtained by repeated simulations on the five-spin TFIM at $h/J=0.5$ with one ancilla qubit. (a) The number of successful trials of different classical optimizers. (b) The average iterations when a different learning rate $\eta$ is applied in SPSA, with fixed perturbation of $\epsilon = 0.01$. (c) The average iterations when different perturbation $\epsilon$ is chosen. The learning rate is fixed to $\eta = 0.2$. The red crosses indicate the failed optimizations.}
	\label{Fig6}
\end{figure} 

\begin{figure*}[t]
	\centering
	\includegraphics[width=0.89\textwidth]{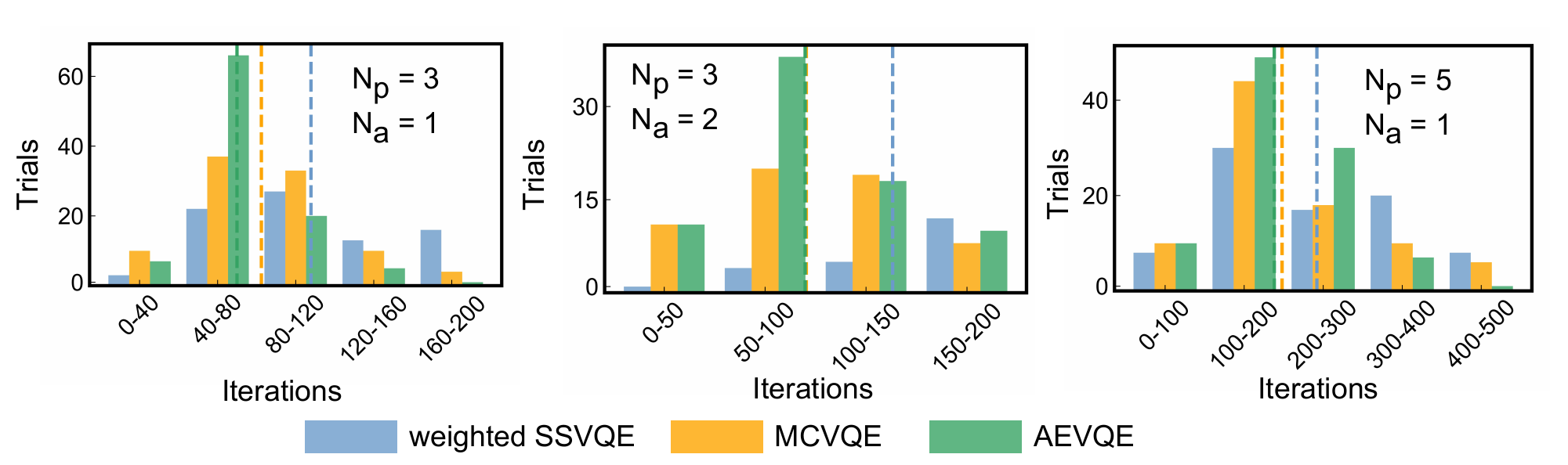}
	\caption{The distribution of the required number of iterations when the loss deviation from the exact energy falls below 0.05 before reaching the maximum iteration limit (200 for $N_p$ =3 and 500 for $N_p =5$). The blue, orange, and green bars represent the the weighted SSVQE, MCVQE, and AEVQE, the dashed lines represents the average iteration. The simulations are carried out with three TFIM cases: $N_p =3, N_a =1$, $N_p=3, N_a =2$, and $N_p =5, N_a =1$, at $h/J=0.5$, respectively. }
	\label{Fig7}
\end{figure*} 

We further discuss the optimization efficiency of AEVQE by adjusting the type of classical optimizers, and the hyperparameters of the SPSA optimizer. First, we calculate two lowest eigenenergies in a five-spin TFIM using SPSA, Batch Gradient Descent (BGD) \cite{ruder2016overview}, Powell \cite{powell1970survey}, and the Genetic Algorithm (GA) \cite{grefenstette1993genetic}. The first two optimizers are gradient-based, while the latter two are gradient-free. We count the number of successful trials among 100 repeated simulations. The results in Fig. \ref{Fig6}(a) show that the choice of classical optimizers is flexible, and gradient-based optimizers generally yield a higher success probability.

The effects of the learning rate and perturbation on the  SPSA optimizer are shown in Figs. \ref{Fig6}(b) and (c). For the five-spin TFIM, the convergence requires fewer iterations when the learning rate $\eta$ is in  the range of $0.2-0.3$, and the perturbation $\epsilon$ is in the range of $0.05-0.15$. With improper parameters, such as a learning rate of 0.5 or a perturbation of 0.01, the optimization fails. The learning rate influences the optimization efficiency more than the perturbation. The learning rate has more influence on the optimization efficiency than the perturbation. 

\section{Comparison with ancilla-free algorithms}

In this section, we review two representative ancilla-free algorithms, i,e. weighted SSVQE and MCVQE, and compare them with  AEVQE in both optimization and diagonalization stages, respectively.   

\subsection {Optimization stage}
We first compare the shots budget of different algorithms in the optimization stage, as the optimization stage requires orders of magnitude more shots than the diagonalization stage. The shot budget follows a triple product: shots per circuit  $\times$ circuits per iteration $\times$ number of iterations.

\subsubsection{Shots per circuit} 

In the presence of sampling error, the standard deviation $\epsilon_0$ of an expectation value in AEVQE is proportional to $\sqrt{1/M_0}$, where $M_0$ is the number of shots per measurement. \cite{giovannetti2011advances}. For weighted SSVQE, when each quantum circuit with weight $w_i$ is allocated $M_{1i}$ shots, the combined standard deviation is calculated as 

\begin{equation}
 \epsilon_1 = \sqrt{\sum_{i=0}^{K-1}\epsilon_{1i}^{2}}\sim\sqrt{\sum_{i=0}^{K-1}\frac{w_i^2}{M_{1i}}}.
\end{equation}
If we consider $\epsilon_1 = \epsilon_0$ and each component contributes equally to the combined standard deviation, we have $\epsilon_{1i}^2 = \epsilon_0^2/K$, the required shots per circuit $M_{1i} = w_{i}^{2}KM_{0}$. For MCVQE, as a special case of the weighted SSVQE with uniform weights of $w_i = 1/K$, the required shots per circuit satisfies $M_{1i} = M_0/K$. 

\subsubsection{Circuits per iteration}

In AEVQE, $K$ variational states are simultaneously prepared by entangling $N_a$ ancilla qubits with $N_a$ physical qubits ($N_a = \log_2K$). The loss function can be calculated by measuring only one quantum circuit. In contrast, in weighted SSVQE and MCVQE, the physical qubits are initialized to $K$ orthogonal states successively, and the calculation of the loss function requires measuring $K$ quantum circuits. As a result, the circuit-per-iteration count is reduced by a factor of $K$ in AEVQE.

\subsubsection{Number of iterations}  
We compare the number of iterations of the three algorithms in three TFIM cases: $N_p=3, N_a=1$, $N_p =3, N_a=2$, and $N_p=5, N_a=1$. For each algorithm, the optimization stage is repeated by 100 trials with random initial values.  We define an optimization successful if the loss deviation from the exact energy falls below 0.05 before reaching the maximum iteration limit (200 for $N_p$ =3 and 500 for $N_p =5$). We record the required iterations of successful trials, and exclude failed trials from statistics. The distribution of recorded numbers of iteration are shown in Fig. \ref{Fig7}. 

In all three cases, the weighted SSVQE exhibits slower convergence, requiring  $107\pm44$, $138\pm40$, and $242\pm111$ iterations. In contrast, AEVQE  converges in $69\pm28$, $94\pm43$, and $188\pm77$ iterations, while MCVQE requires  $82\pm36$, $95\pm44$, and $198\pm103$ iterations.  Furthermore, when targeting four low-lying energy levels, AEVQE and MCVQE obtain more success trials (77 and 58 successful trials, respectively) than the weighted SSVQE algorithm, which only succeeds in 22 trials. 

An intuitive explanation for this performance gap is that for AEVQE or MCVQE, the global minimum is reached when the variational state is optimized to the subspace spanned by the low-lying eigenstates, while  the weighted SSVQE demands excat mapping of each initial state to the corresponding eigenstate, which imposes stricter optimization constrains.

\subsection{Diagonalization stage} 

For weighted SSVQE, eigenstates can be prepared by implementing the variational circuit with the optimal variational parameters, no additional diagonalization is required. In contrast, as the MCVQE and AEVQE only converge to the $K-$ dimensional subspace, a subsequent diagnolization is required to exact eigenenergies. For MCVQE, the matrix elements for the Hamiltonian  in the $K$ subspace are given by
\begin{equation*}
    H_{\text{sub}}[m,m] = \langle\psi_{m}| U^{\dagger}(\bm{\theta})HU(\bm{\theta})|\psi_m\rangle, \\
\end{equation*}
\begin{eqnarray}
    H_{\text{sub}}[m,n] =\frac{\langle\psi^{+}_{m,n}|U(\bm{\theta})^{\dagger}HU(\bm{\theta})|\psi^{+}_{m,n}\rangle}{4} \\
    -\frac{\langle\psi^{-}_{m,n}| U(\bm{\theta})^{\dagger}HU(\bm{\theta})|\psi^{-}_{m,n}\rangle}{4} \nonumber.
\end{eqnarray}
where $|\psi^{\pm}_{m,n}\rangle$ represents the $|\psi_{m}\rangle\pm|\psi_n\rangle$, respectively~\cite{parrish2019quantum}. The eigenenergies are then obtained through classical diagnolization of $H_{\text{sub}}$.

We note that, MCVQE can also prepare the target eigenstates by implementing the diagonaling unitary operation on the system qubits before the variational circuit with optimized parameters, which is equivalent to AEVQE due to the Choi-Jamiołkowski isomorphism \cite{jamiolkowski1972linear,choi1975completely}. As a result, AEVQE can be applied as an alternative to MCVQE, reducing the circuit depth at the cost of additional ancilla qubit, thus may increase the accuracy when the main error originates from the decoherence of the physical qubits.

\section{Conclusion and outlook} 

We have employed the AEVQE to simultaneously calculate multiple eigenenergies and eigenstates of the $\text{H}_2$ molecule as well as the transverse field Ising models on a superconducting quantum processor.  Using these results, we have plotted the potential energy curve of the $\text{H}_2$ molecule, and measured the average absolute magnetization to indicate the phase transition in the TFIMs.  Moreover, we have carried out simulations to analyze the effects of system size and classical optimizers on the optimization efficiency, and compared our algorithm with the previous ancilla-free algorithms. Our results demonstrate the  feasibility of AEVQE on the current noisy quantum hardware. 

Our results also highlight the potential challenges of AEVQE. As the number of ancilla qubits increases,  ancilla-physical entangelement becomes more challenging if all qubits are arranged in a one-dimensional chain. This problem may be alleviated by using qubits that are arranged with a ladder-shaped architecture. Furthormore, implementing the exact diagonalization unitary for  Hamiltonian in the subspaces  can be demanding when targeting a large number of low-lying energy levels.  A possible solution is to introduce another optimization stage to realize approximate diagonalization. Finally, similar to other VQAs, as the system size increases, the barren plateau phenomenon becomes more severe. Developing strategies to avoid the barren plateaus in experiments requires further explorations. 

\section*{Acknowledgments}

We acknowledge supports from the National Natural Science Foundation of China (Grants Nos. 92365206,12404560), the Innovation Program for Quantum Science and Technology (Grant No. 2021ZD0301802).

\section*{Data availability}

The data generated in this study have been deposited in the zenodo database \cite{zenodo}.


\begin{thebibliography}{70}%
	\makeatletter
	\providecommand \@ifxundefined [1]{%
		\@ifx{#1\undefined}
	}%
	\providecommand \@ifnum [1]{%
		\ifnum #1\expandafter \@firstoftwo
		\else \expandafter \@secondoftwo
		\fi
	}%
	\providecommand \@ifx [1]{%
		\ifx #1\expandafter \@firstoftwo
		\else \expandafter \@secondoftwo
		\fi
	}%
	\providecommand \natexlab [1]{#1}%
	\providecommand \enquote  [1]{``#1''}%
	\providecommand \bibnamefont  [1]{#1}%
	\providecommand \bibfnamefont [1]{#1}%
	\providecommand \citenamefont [1]{#1}%
	\providecommand \href@noop [0]{\@secondoftwo}%
	\providecommand \href [0]{\begingroup \@sanitize@url \@href}%
	\providecommand \@href[1]{\@@startlink{#1}\@@href}%
	\providecommand \@@href[1]{\endgroup#1\@@endlink}%
	\providecommand \@sanitize@url [0]{\catcode `\\12\catcode `\$12\catcode
		`\&12\catcode `\#12\catcode `\^12\catcode `\_12\catcode `\%12\relax}%
	\providecommand \@@startlink[1]{}%
	\providecommand \@@endlink[0]{}%
	\providecommand \url  [0]{\begingroup\@sanitize@url \@url }%
	\providecommand \@url [1]{\endgroup\@href {#1}{\urlprefix }}%
	\providecommand \urlprefix  [0]{URL }%
	\providecommand \Eprint [0]{\href }%
	\providecommand \doibase [0]{https://doi.org/}%
	\providecommand \selectlanguage [0]{\@gobble}%
	\providecommand \bibinfo  [0]{\@secondoftwo}%
	\providecommand \bibfield  [0]{\@secondoftwo}%
	\providecommand \translation [1]{[#1]}%
	\providecommand \BibitemOpen [0]{}%
	\providecommand \bibitemStop [0]{}%
	\providecommand \bibitemNoStop [0]{.\EOS\space}%
	\providecommand \EOS [0]{\spacefactor3000\relax}%
	\providecommand \BibitemShut  [1]{\csname bibitem#1\endcsname}%
	\let\auto@bib@innerbib\@empty
	\bibitem [{\citenamefont {Lloyd}(1996)}]{lloyd1996universal}%
	\BibitemOpen
	\bibfield  {author} {\bibinfo {author} {\bibfnamefont {S.}~\bibnamefont
			{Lloyd}},\ }\bibfield  {title} {\bibinfo {title} {Universal quantum
			simulators},\ }\href
	{https://www.science.org/doi/10.1126/science.273.5278.1073} {\bibfield
		{journal} {\bibinfo  {journal} {Science}\ }\textbf {\bibinfo {volume}
			{273}},\ \bibinfo {pages} {1073} (\bibinfo {year} {1996})}\BibitemShut
	{NoStop}%
	\bibitem [{\citenamefont {Georgescu}\ \emph {et~al.}(2014)\citenamefont
		{Georgescu}, \citenamefont {Ashhab},\ and\ \citenamefont
		{Nori}}]{georgescu2014quantum}%
	\BibitemOpen
	\bibfield  {author} {\bibinfo {author} {\bibfnamefont {I.~M.}\ \bibnamefont
			{Georgescu}}, \bibinfo {author} {\bibfnamefont {S.}~\bibnamefont {Ashhab}},\
		and\ \bibinfo {author} {\bibfnamefont {F.}~\bibnamefont {Nori}},\ }\bibfield
	{title} {\bibinfo {title} {Quantum simulation},\ }\href
	{https://journals.aps.org/rmp/abstract/10.1103/RevModPhys.86.153} {\bibfield
		{journal} {\bibinfo  {journal} {Rev. Mod. Phys.}\ }\textbf {\bibinfo {volume}
			{86}},\ \bibinfo {pages} {153} (\bibinfo {year} {2014})}\BibitemShut
	{NoStop}%
	\bibitem [{\citenamefont {Arute}\ \emph {et~al.}(2019)\citenamefont {Arute},
		\citenamefont {Arya}, \citenamefont {Babbush}, \citenamefont {Bacon},
		\citenamefont {Bardin}, \citenamefont {Barends}, \citenamefont {Biswas},
		\citenamefont {Boixo}, \citenamefont {Brandao}, \citenamefont {Buell} \emph
		{et~al.}}]{arute2019quantum}%
	\BibitemOpen
	\bibfield  {author} {\bibinfo {author} {\bibfnamefont {F.}~\bibnamefont
			{Arute}}, \bibinfo {author} {\bibfnamefont {K.}~\bibnamefont {Arya}},
		\bibinfo {author} {\bibfnamefont {R.}~\bibnamefont {Babbush}}, \bibinfo
		{author} {\bibfnamefont {D.}~\bibnamefont {Bacon}}, \bibinfo {author}
		{\bibfnamefont {J.~C.}\ \bibnamefont {Bardin}}, \bibinfo {author}
		{\bibfnamefont {R.}~\bibnamefont {Barends}}, \bibinfo {author} {\bibfnamefont
			{R.}~\bibnamefont {Biswas}}, \bibinfo {author} {\bibfnamefont
			{S.}~\bibnamefont {Boixo}}, \bibinfo {author} {\bibfnamefont {F.~G.}\
			\bibnamefont {Brandao}}, \bibinfo {author} {\bibfnamefont {D.~A.}\
			\bibnamefont {Buell}}, \emph {et~al.},\ }\bibfield  {title} {\bibinfo {title}
		{Quantum supremacy using a programmable superconducting processor},\ }\href
	{https://doi.org/https://doi.org/10.1038/s41586-019-1666-5} {\bibfield
		{journal} {\bibinfo  {journal} {Nature}\ }\textbf {\bibinfo {volume} {574}},\
		\bibinfo {pages} {505} (\bibinfo {year} {2019})}\BibitemShut {NoStop}%
	\bibitem [{\citenamefont {Zhong}\ \emph {et~al.}(2020)\citenamefont {Zhong},
		\citenamefont {Wang}, \citenamefont {Deng}, \citenamefont {Chen},
		\citenamefont {Peng}, \citenamefont {Luo}, \citenamefont {Qin}, \citenamefont
		{Wu}, \citenamefont {Ding}, \citenamefont {Hu} \emph
		{et~al.}}]{zhong2020quantum}%
	\BibitemOpen
	\bibfield  {author} {\bibinfo {author} {\bibfnamefont {H.-S.}\ \bibnamefont
			{Zhong}}, \bibinfo {author} {\bibfnamefont {H.}~\bibnamefont {Wang}},
		\bibinfo {author} {\bibfnamefont {Y.-H.}\ \bibnamefont {Deng}}, \bibinfo
		{author} {\bibfnamefont {M.-C.}\ \bibnamefont {Chen}}, \bibinfo {author}
		{\bibfnamefont {L.-C.}\ \bibnamefont {Peng}}, \bibinfo {author}
		{\bibfnamefont {Y.-H.}\ \bibnamefont {Luo}}, \bibinfo {author} {\bibfnamefont
			{J.}~\bibnamefont {Qin}}, \bibinfo {author} {\bibfnamefont {D.}~\bibnamefont
			{Wu}}, \bibinfo {author} {\bibfnamefont {X.}~\bibnamefont {Ding}}, \bibinfo
		{author} {\bibfnamefont {Y.}~\bibnamefont {Hu}}, \emph {et~al.},\ }\bibfield
	{title} {\bibinfo {title} {Quantum computational advantage using photons},\
	}\href {https://doi.org/DOI:10.1126/science.abe8770} {\bibfield  {journal}
		{\bibinfo  {journal} {Science}\ }\textbf {\bibinfo {volume} {370}},\ \bibinfo
		{pages} {1460} (\bibinfo {year} {2020})}\BibitemShut {NoStop}%
	\bibitem [{\citenamefont {Madsen}\ \emph {et~al.}(2022)\citenamefont {Madsen},
		\citenamefont {Laudenbach}, \citenamefont {Askarani}, \citenamefont
		{Rortais}, \citenamefont {Vincent}, \citenamefont {Bulmer}, \citenamefont
		{Miatto}, \citenamefont {Neuhaus}, \citenamefont {Helt}, \citenamefont
		{Collins} \emph {et~al.}}]{madsen2022quantum}%
	\BibitemOpen
	\bibfield  {author} {\bibinfo {author} {\bibfnamefont {L.~S.}\ \bibnamefont
			{Madsen}}, \bibinfo {author} {\bibfnamefont {F.}~\bibnamefont {Laudenbach}},
		\bibinfo {author} {\bibfnamefont {M.~F.}\ \bibnamefont {Askarani}}, \bibinfo
		{author} {\bibfnamefont {F.}~\bibnamefont {Rortais}}, \bibinfo {author}
		{\bibfnamefont {T.}~\bibnamefont {Vincent}}, \bibinfo {author} {\bibfnamefont
			{J.~F.}\ \bibnamefont {Bulmer}}, \bibinfo {author} {\bibfnamefont {F.~M.}\
			\bibnamefont {Miatto}}, \bibinfo {author} {\bibfnamefont {L.}~\bibnamefont
			{Neuhaus}}, \bibinfo {author} {\bibfnamefont {L.~G.}\ \bibnamefont {Helt}},
		\bibinfo {author} {\bibfnamefont {M.~J.}\ \bibnamefont {Collins}}, \emph
		{et~al.},\ }\bibfield  {title} {\bibinfo {title} {Quantum computational
			advantage with a programmable photonic processor},\ }\href
	{https://doi.org/https://doi.org/10.1038/s41586-022-04725-x} {\bibfield
		{journal} {\bibinfo  {journal} {Nature}\ }\textbf {\bibinfo {volume} {606}},\
		\bibinfo {pages} {75} (\bibinfo {year} {2022})}\BibitemShut {NoStop}%
	\bibitem [{\citenamefont {Shor}(1997)}]{shor1999polynomial}%
	\BibitemOpen
	\bibfield  {author} {\bibinfo {author} {\bibfnamefont {P.~W.}\ \bibnamefont
			{Shor}},\ }\bibfield  {title} {\bibinfo {title} {Polynomial-time algorithms
			for prime factorization and discrete logarithms on a quantum computer},\
	}\href {https://doi.org/10.1137/S0097539795293172} {\bibfield  {journal}
		{\bibinfo  {journal} {SIAM J. Comput.,}\ }\textbf {\bibinfo {volume} {26}},\
		\bibinfo {pages} {1484} (\bibinfo {year} {1997})}\BibitemShut {NoStop}%
	\bibitem [{\citenamefont {Aspuru-Guzik}\ \emph {et~al.}(2005)\citenamefont
		{Aspuru-Guzik}, \citenamefont {Dutoi}, \citenamefont {Love},\ and\
		\citenamefont {Head-Gordon}}]{aspuru2005simulated}%
	\BibitemOpen
	\bibfield  {author} {\bibinfo {author} {\bibfnamefont {A.}~\bibnamefont
			{Aspuru-Guzik}}, \bibinfo {author} {\bibfnamefont {A.~D.}\ \bibnamefont
			{Dutoi}}, \bibinfo {author} {\bibfnamefont {P.~J.}\ \bibnamefont {Love}},\
		and\ \bibinfo {author} {\bibfnamefont {M.}~\bibnamefont {Head-Gordon}},\
	}\bibfield  {title} {\bibinfo {title} {Simulated quantum computation of
			molecular energies},\ }\href {https://doi.org/DOI:10.1126/science.1113479}
	{\bibfield  {journal} {\bibinfo  {journal} {Science}\ }\textbf {\bibinfo
			{volume} {309}},\ \bibinfo {pages} {1704} (\bibinfo {year}
		{2005})}\BibitemShut {NoStop}%
	\bibitem [{\citenamefont {Yuan}(2020)}]{yuan2020quantum}%
	\BibitemOpen
	\bibfield  {author} {\bibinfo {author} {\bibfnamefont {X.}~\bibnamefont
			{Yuan}},\ }\bibfield  {title} {\bibinfo {title} {A quantum-computing
			advantage for chemistry},\ }\href
	{https://doi.org/DOI:10.1126/science.abd3880} {\bibfield  {journal} {\bibinfo
			{journal} {Science}\ }\textbf {\bibinfo {volume} {369}},\ \bibinfo {pages}
		{1054} (\bibinfo {year} {2020})}\BibitemShut {NoStop}%
	\bibitem [{\citenamefont {McArdle}\ \emph {et~al.}(2020)\citenamefont
		{McArdle}, \citenamefont {Endo}, \citenamefont {Aspuru-Guzik}, \citenamefont
		{Benjamin},\ and\ \citenamefont {Yuan}}]{mcardle2020quantum}%
	\BibitemOpen
	\bibfield  {author} {\bibinfo {author} {\bibfnamefont {S.}~\bibnamefont
			{McArdle}}, \bibinfo {author} {\bibfnamefont {S.}~\bibnamefont {Endo}},
		\bibinfo {author} {\bibfnamefont {A.}~\bibnamefont {Aspuru-Guzik}}, \bibinfo
		{author} {\bibfnamefont {S.~C.}\ \bibnamefont {Benjamin}},\ and\ \bibinfo
		{author} {\bibfnamefont {X.}~\bibnamefont {Yuan}},\ }\bibfield  {title}
	{\bibinfo {title} {Quantum computational chemistry},\ }\href
	{https://doi.org/DOI: https://doi.org/10.1103/RevModPhys.92.015003}
	{\bibfield  {journal} {\bibinfo  {journal} {Rev. Mod. Phys.}\ }\textbf
		{\bibinfo {volume} {92}},\ \bibinfo {pages} {015003} (\bibinfo {year}
		{2020})}\BibitemShut {NoStop}%
	\bibitem [{\citenamefont {Bauer}\ \emph {et~al.}(2023)\citenamefont {Bauer},
		\citenamefont {Davoudi}, \citenamefont {Balantekin}, \citenamefont
		{Bhattacharya}, \citenamefont {Carena}, \citenamefont {De~Jong},
		\citenamefont {Draper}, \citenamefont {El-Khadra}, \citenamefont {Gemelke},
		\citenamefont {Hanada} \emph {et~al.}}]{bauer2023quantum}%
	\BibitemOpen
	\bibfield  {author} {\bibinfo {author} {\bibfnamefont {C.~W.}\ \bibnamefont
			{Bauer}}, \bibinfo {author} {\bibfnamefont {Z.}~\bibnamefont {Davoudi}},
		\bibinfo {author} {\bibfnamefont {A.~B.}\ \bibnamefont {Balantekin}},
		\bibinfo {author} {\bibfnamefont {T.}~\bibnamefont {Bhattacharya}}, \bibinfo
		{author} {\bibfnamefont {M.}~\bibnamefont {Carena}}, \bibinfo {author}
		{\bibfnamefont {W.~A.}\ \bibnamefont {De~Jong}}, \bibinfo {author}
		{\bibfnamefont {P.}~\bibnamefont {Draper}}, \bibinfo {author} {\bibfnamefont
			{A.}~\bibnamefont {El-Khadra}}, \bibinfo {author} {\bibfnamefont
			{N.}~\bibnamefont {Gemelke}}, \bibinfo {author} {\bibfnamefont
			{M.}~\bibnamefont {Hanada}}, \emph {et~al.},\ }\bibfield  {title} {\bibinfo
		{title} {Quantum simulation for high-energy physics},\ }\href
	{https://doi.org/https://doi.org/10.1103/PRXQuantum.4.027001} {\bibfield
		{journal} {\bibinfo  {journal} {PRX Quantum}\ }\textbf {\bibinfo {volume}
			{4}},\ \bibinfo {pages} {027001} (\bibinfo {year} {2023})}\BibitemShut
	{NoStop}%
	\bibitem [{\citenamefont {Gerritsma}\ \emph {et~al.}(2010)\citenamefont
		{Gerritsma}, \citenamefont {Kirchmair}, \citenamefont {Z{\"a}hringer},
		\citenamefont {Solano}, \citenamefont {Blatt},\ and\ \citenamefont
		{Roos}}]{gerritsma2010quantum}%
	\BibitemOpen
	\bibfield  {author} {\bibinfo {author} {\bibfnamefont {R.}~\bibnamefont
			{Gerritsma}}, \bibinfo {author} {\bibfnamefont {G.}~\bibnamefont
			{Kirchmair}}, \bibinfo {author} {\bibfnamefont {F.}~\bibnamefont
			{Z{\"a}hringer}}, \bibinfo {author} {\bibfnamefont {E.}~\bibnamefont
			{Solano}}, \bibinfo {author} {\bibfnamefont {R.}~\bibnamefont {Blatt}},\ and\
		\bibinfo {author} {\bibfnamefont {C.}~\bibnamefont {Roos}},\ }\bibfield
	{title} {\bibinfo {title} {Quantum simulation of the dirac equation},\ }\href
	{https://doi.org/https://doi.org/10.1038/nature08688} {\bibfield  {journal}
		{\bibinfo  {journal} {Nature}\ }\textbf {\bibinfo {volume} {463}},\ \bibinfo
		{pages} {68} (\bibinfo {year} {2010})}\BibitemShut {NoStop}%
	\bibitem [{\citenamefont {Bloch}\ \emph {et~al.}(2012)\citenamefont {Bloch},
		\citenamefont {Dalibard},\ and\ \citenamefont
		{Nascimbene}}]{bloch2012quantum}%
	\BibitemOpen
	\bibfield  {author} {\bibinfo {author} {\bibfnamefont {I.}~\bibnamefont
			{Bloch}}, \bibinfo {author} {\bibfnamefont {J.}~\bibnamefont {Dalibard}},\
		and\ \bibinfo {author} {\bibfnamefont {S.}~\bibnamefont {Nascimbene}},\
	}\bibfield  {title} {\bibinfo {title} {Quantum simulations with ultracold
			quantum gases},\ }\href {https://doi.org/https://doi.org/10.1038/nphys2259}
	{\bibfield  {journal} {\bibinfo  {journal} {Nat. Phys.}\ }\textbf {\bibinfo
			{volume} {8}},\ \bibinfo {pages} {267} (\bibinfo {year} {2012})}\BibitemShut
	{NoStop}%
	\bibitem [{\citenamefont {Jordan}\ and\ \citenamefont
		{Mitchell}(2015)}]{jordan2015machine}%
	\BibitemOpen
	\bibfield  {author} {\bibinfo {author} {\bibfnamefont {M.~I.}\ \bibnamefont
			{Jordan}}\ and\ \bibinfo {author} {\bibfnamefont {T.~M.}\ \bibnamefont
			{Mitchell}},\ }\bibfield  {title} {\bibinfo {title} {Machine learning:
			Trends, perspectives, and prospects},\ }\href {https://doi.org/DOI:
		10.1126/science.aaa8415} {\bibfield  {journal} {\bibinfo  {journal}
			{Science}\ }\textbf {\bibinfo {volume} {349}},\ \bibinfo {pages} {255}
		(\bibinfo {year} {2015})}\BibitemShut {NoStop}%
	\bibitem [{\citenamefont {Carleo}\ \emph {et~al.}(2019)\citenamefont {Carleo},
		\citenamefont {Cirac}, \citenamefont {Cranmer}, \citenamefont {Daudet},
		\citenamefont {Schuld}, \citenamefont {Tishby}, \citenamefont
		{Vogt-Maranto},\ and\ \citenamefont {Zdeborov{\'a}}}]{carleo2019machine}%
	\BibitemOpen
	\bibfield  {author} {\bibinfo {author} {\bibfnamefont {G.}~\bibnamefont
			{Carleo}}, \bibinfo {author} {\bibfnamefont {I.}~\bibnamefont {Cirac}},
		\bibinfo {author} {\bibfnamefont {K.}~\bibnamefont {Cranmer}}, \bibinfo
		{author} {\bibfnamefont {L.}~\bibnamefont {Daudet}}, \bibinfo {author}
		{\bibfnamefont {M.}~\bibnamefont {Schuld}}, \bibinfo {author} {\bibfnamefont
			{N.}~\bibnamefont {Tishby}}, \bibinfo {author} {\bibfnamefont
			{L.}~\bibnamefont {Vogt-Maranto}},\ and\ \bibinfo {author} {\bibfnamefont
			{L.}~\bibnamefont {Zdeborov{\'a}}},\ }\bibfield  {title} {\bibinfo {title}
		{Machine learning and the physical sciences},\ }\href
	{https://doi.org/https://doi.org/10.1103/RevModPhys.91.045002} {\bibfield
		{journal} {\bibinfo  {journal} {Rev. Mod. Phys.}\ }\textbf {\bibinfo {volume}
			{91}},\ \bibinfo {pages} {045002} (\bibinfo {year} {2019})}\BibitemShut
	{NoStop}%
	\bibitem [{\citenamefont {Deng}\ \emph {et~al.}(2017)\citenamefont {Deng},
		\citenamefont {Li},\ and\ \citenamefont {Das~Sarma}}]{deng2017machine}%
	\BibitemOpen
	\bibfield  {author} {\bibinfo {author} {\bibfnamefont {D.-L.}\ \bibnamefont
			{Deng}}, \bibinfo {author} {\bibfnamefont {X.}~\bibnamefont {Li}},\ and\
		\bibinfo {author} {\bibfnamefont {S.}~\bibnamefont {Das~Sarma}},\ }\bibfield
	{title} {\bibinfo {title} {Machine learning topological states},\ }\href
	{https://doi.org/DOI: https://doi.org/10.1103/PhysRevB.96.195145} {\bibfield
		{journal} {\bibinfo  {journal} {Phys. Rev. B}\ }\textbf {\bibinfo {volume}
			{96}},\ \bibinfo {pages} {195145} (\bibinfo {year} {2017})}\BibitemShut
	{NoStop}%
	\bibitem [{\citenamefont {Bausch}\ \emph {et~al.}(2024)\citenamefont {Bausch},
		\citenamefont {Senior}, \citenamefont {Heras}, \citenamefont {Edlich},
		\citenamefont {Davies}, \citenamefont {Newman}, \citenamefont {Jones},
		\citenamefont {Satzinger}, \citenamefont {Niu}, \citenamefont {Blackwell}
		\emph {et~al.}}]{bausch2024learning}%
	\BibitemOpen
	\bibfield  {author} {\bibinfo {author} {\bibfnamefont {J.}~\bibnamefont
			{Bausch}}, \bibinfo {author} {\bibfnamefont {A.~W.}\ \bibnamefont {Senior}},
		\bibinfo {author} {\bibfnamefont {F.~J.}\ \bibnamefont {Heras}}, \bibinfo
		{author} {\bibfnamefont {T.}~\bibnamefont {Edlich}}, \bibinfo {author}
		{\bibfnamefont {A.}~\bibnamefont {Davies}}, \bibinfo {author} {\bibfnamefont
			{M.}~\bibnamefont {Newman}}, \bibinfo {author} {\bibfnamefont
			{C.}~\bibnamefont {Jones}}, \bibinfo {author} {\bibfnamefont
			{K.}~\bibnamefont {Satzinger}}, \bibinfo {author} {\bibfnamefont {M.~Y.}\
			\bibnamefont {Niu}}, \bibinfo {author} {\bibfnamefont {S.}~\bibnamefont
			{Blackwell}}, \emph {et~al.},\ }\bibfield  {title} {\bibinfo {title}
		{Learning high-accuracy error decoding for quantum processors},\ }\href
	{https://doi.org/https://doi.org/10.1038/s41586-024-08148-8} {\bibfield
		{journal} {\bibinfo  {journal} {Nature}\ }\textbf {\bibinfo {volume} {635}},\
		\bibinfo {pages} {834} (\bibinfo {year} {2024})}\BibitemShut {NoStop}%
	\bibitem [{\citenamefont {Alexeev}\ \emph {et~al.}(2024)\citenamefont
		{Alexeev}, \citenamefont {Farag}, \citenamefont {Patti}, \citenamefont
		{Wolf}, \citenamefont {Ares}, \citenamefont {Aspuru-Guzik}, \citenamefont
		{Benjamin}, \citenamefont {Cai}, \citenamefont {Chandani}, \citenamefont
		{Fedele} \emph {et~al.}}]{alexeev2024artificial}%
	\BibitemOpen
	\bibfield  {author} {\bibinfo {author} {\bibfnamefont {Y.}~\bibnamefont
			{Alexeev}}, \bibinfo {author} {\bibfnamefont {M.~H.}\ \bibnamefont {Farag}},
		\bibinfo {author} {\bibfnamefont {T.~L.}\ \bibnamefont {Patti}}, \bibinfo
		{author} {\bibfnamefont {M.~E.}\ \bibnamefont {Wolf}}, \bibinfo {author}
		{\bibfnamefont {N.}~\bibnamefont {Ares}}, \bibinfo {author} {\bibfnamefont
			{A.}~\bibnamefont {Aspuru-Guzik}}, \bibinfo {author} {\bibfnamefont {S.~C.}\
			\bibnamefont {Benjamin}}, \bibinfo {author} {\bibfnamefont {Z.}~\bibnamefont
			{Cai}}, \bibinfo {author} {\bibfnamefont {Z.}~\bibnamefont {Chandani}},
		\bibinfo {author} {\bibfnamefont {F.}~\bibnamefont {Fedele}}, \emph
		{et~al.},\ }\bibfield  {title} {\bibinfo {title} {Artificial intelligence for
			quantum computing},\ }\href@noop {} {\bibfield  {journal} {\bibinfo
			{journal} {preprint arXiv:2411.09131}\ } (\bibinfo {year}
		{2024})}\BibitemShut {NoStop}%
	\bibitem [{\citenamefont {Koutromanos}\ \emph {et~al.}(2024)\citenamefont
		{Koutromanos}, \citenamefont {Stefanatos},\ and\ \citenamefont
		{Paspalakis}}]{koutromanos2024control}%
	\BibitemOpen
	\bibfield  {author} {\bibinfo {author} {\bibfnamefont {D.}~\bibnamefont
			{Koutromanos}}, \bibinfo {author} {\bibfnamefont {D.}~\bibnamefont
			{Stefanatos}},\ and\ \bibinfo {author} {\bibfnamefont {E.}~\bibnamefont
			{Paspalakis}},\ }\bibfield  {title} {\bibinfo {title} {Control of qubit
			dynamics using reinforcement learning},\ }\href
	{https://doi.org/https://doi.org/10.3390/info15050272} {\bibfield  {journal}
		{\bibinfo  {journal} {Information}\ }\textbf {\bibinfo {volume} {15}},\
		\bibinfo {pages} {272} (\bibinfo {year} {2024})}\BibitemShut {NoStop}%
	\bibitem [{\citenamefont {Cerezo}\ \emph {et~al.}(2021)\citenamefont {Cerezo},
		\citenamefont {Arrasmith}, \citenamefont {Babbush}, \citenamefont {Benjamin},
		\citenamefont {Endo}, \citenamefont {Fujii}, \citenamefont {McClean},
		\citenamefont {Mitarai}, \citenamefont {Yuan}, \citenamefont {Cincio} \emph
		{et~al.}}]{cerezo2021variational}%
	\BibitemOpen
	\bibfield  {author} {\bibinfo {author} {\bibfnamefont {M.}~\bibnamefont
			{Cerezo}}, \bibinfo {author} {\bibfnamefont {A.}~\bibnamefont {Arrasmith}},
		\bibinfo {author} {\bibfnamefont {R.}~\bibnamefont {Babbush}}, \bibinfo
		{author} {\bibfnamefont {S.~C.}\ \bibnamefont {Benjamin}}, \bibinfo {author}
		{\bibfnamefont {S.}~\bibnamefont {Endo}}, \bibinfo {author} {\bibfnamefont
			{K.}~\bibnamefont {Fujii}}, \bibinfo {author} {\bibfnamefont {J.~R.}\
			\bibnamefont {McClean}}, \bibinfo {author} {\bibfnamefont {K.}~\bibnamefont
			{Mitarai}}, \bibinfo {author} {\bibfnamefont {X.}~\bibnamefont {Yuan}},
		\bibinfo {author} {\bibfnamefont {L.}~\bibnamefont {Cincio}}, \emph
		{et~al.},\ }\bibfield  {title} {\bibinfo {title} {Variational quantum
			algorithms},\ }\href
	{https://doi.org/https://doi.org/10.1038/s42254-021-00348-9} {\bibfield
		{journal} {\bibinfo  {journal} {Nat. Rev. Phys.}\ }\textbf {\bibinfo {volume}
			{3}},\ \bibinfo {pages} {625} (\bibinfo {year} {2021})}\BibitemShut {NoStop}%
	\bibitem [{\citenamefont {Bharti}\ \emph {et~al.}(2022)\citenamefont {Bharti},
		\citenamefont {Cervera-Lierta}, \citenamefont {Kyaw}, \citenamefont {Haug},
		\citenamefont {Alperin-Lea}, \citenamefont {Anand}, \citenamefont {Degroote},
		\citenamefont {Heimonen}, \citenamefont {Kottmann}, \citenamefont {Menke}
		\emph {et~al.}}]{bharti2022noisy}%
	\BibitemOpen
	\bibfield  {author} {\bibinfo {author} {\bibfnamefont {K.}~\bibnamefont
			{Bharti}}, \bibinfo {author} {\bibfnamefont {A.}~\bibnamefont
			{Cervera-Lierta}}, \bibinfo {author} {\bibfnamefont {T.~H.}\ \bibnamefont
			{Kyaw}}, \bibinfo {author} {\bibfnamefont {T.}~\bibnamefont {Haug}}, \bibinfo
		{author} {\bibfnamefont {S.}~\bibnamefont {Alperin-Lea}}, \bibinfo {author}
		{\bibfnamefont {A.}~\bibnamefont {Anand}}, \bibinfo {author} {\bibfnamefont
			{M.}~\bibnamefont {Degroote}}, \bibinfo {author} {\bibfnamefont
			{H.}~\bibnamefont {Heimonen}}, \bibinfo {author} {\bibfnamefont {J.~S.}\
			\bibnamefont {Kottmann}}, \bibinfo {author} {\bibfnamefont {T.}~\bibnamefont
			{Menke}}, \emph {et~al.},\ }\bibfield  {title} {\bibinfo {title} {Noisy
			intermediate-scale quantum algorithms},\ }\href
	{https://doi.org/https://doi.org/10.1103/RevModPhys.94.015004} {\bibfield
		{journal} {\bibinfo  {journal} {Rev. Mod. Phys.}\ }\textbf {\bibinfo {volume}
			{94}},\ \bibinfo {pages} {015004} (\bibinfo {year} {2022})}\BibitemShut
	{NoStop}%
	\bibitem [{\citenamefont {Kandala}\ \emph {et~al.}(2017)\citenamefont
		{Kandala}, \citenamefont {Mezzacapo}, \citenamefont {Temme}, \citenamefont
		{Takita}, \citenamefont {Brink}, \citenamefont {Chow},\ and\ \citenamefont
		{Gambetta}}]{kandala2017hardware}%
	\BibitemOpen
	\bibfield  {author} {\bibinfo {author} {\bibfnamefont {A.}~\bibnamefont
			{Kandala}}, \bibinfo {author} {\bibfnamefont {A.}~\bibnamefont {Mezzacapo}},
		\bibinfo {author} {\bibfnamefont {K.}~\bibnamefont {Temme}}, \bibinfo
		{author} {\bibfnamefont {M.}~\bibnamefont {Takita}}, \bibinfo {author}
		{\bibfnamefont {M.}~\bibnamefont {Brink}}, \bibinfo {author} {\bibfnamefont
			{J.~M.}\ \bibnamefont {Chow}},\ and\ \bibinfo {author} {\bibfnamefont
			{J.~M.}\ \bibnamefont {Gambetta}},\ }\bibfield  {title} {\bibinfo {title}
		{Hardware-efficient variational quantum eigensolver for small molecules and
			quantum magnets},\ }\href
	{https://doi.org/https://doi.org/10.1038/nature23879} {\bibfield  {journal}
		{\bibinfo  {journal} {Nature}\ }\textbf {\bibinfo {volume} {549}},\ \bibinfo
		{pages} {242} (\bibinfo {year} {2017})}\BibitemShut {NoStop}%
	\bibitem [{\citenamefont {Sweke}\ \emph {et~al.}(2020)\citenamefont {Sweke},
		\citenamefont {Wilde}, \citenamefont {Meyer}, \citenamefont {Schuld},
		\citenamefont {F{\"a}hrmann}, \citenamefont {Meynard-Piganeau},\ and\
		\citenamefont {Eisert}}]{sweke2020stochastic}%
	\BibitemOpen
	\bibfield  {author} {\bibinfo {author} {\bibfnamefont {R.}~\bibnamefont
			{Sweke}}, \bibinfo {author} {\bibfnamefont {F.}~\bibnamefont {Wilde}},
		\bibinfo {author} {\bibfnamefont {J.}~\bibnamefont {Meyer}}, \bibinfo
		{author} {\bibfnamefont {M.}~\bibnamefont {Schuld}}, \bibinfo {author}
		{\bibfnamefont {P.~K.}\ \bibnamefont {F{\"a}hrmann}}, \bibinfo {author}
		{\bibfnamefont {B.}~\bibnamefont {Meynard-Piganeau}},\ and\ \bibinfo {author}
		{\bibfnamefont {J.}~\bibnamefont {Eisert}},\ }\bibfield  {title} {\bibinfo
		{title} {Stochastic gradient descent for hybrid quantum-classical
			optimization},\ }\href
	{https://doi.org/https://doi.org/10.22331/q-2020-08-31-314} {\bibfield
		{journal} {\bibinfo  {journal} {Quantum}\ }\textbf {\bibinfo {volume} {4}},\
		\bibinfo {pages} {314} (\bibinfo {year} {2020})}\BibitemShut {NoStop}%
	\bibitem [{\citenamefont {Tang}\ \emph {et~al.}(2021)\citenamefont {Tang},
		\citenamefont {Shkolnikov}, \citenamefont {Barron}, \citenamefont {Grimsley},
		\citenamefont {Mayhall}, \citenamefont {Barnes},\ and\ \citenamefont
		{Economou}}]{tang2021qubit}%
	\BibitemOpen
	\bibfield  {author} {\bibinfo {author} {\bibfnamefont {H.~L.}\ \bibnamefont
			{Tang}}, \bibinfo {author} {\bibfnamefont {V.}~\bibnamefont {Shkolnikov}},
		\bibinfo {author} {\bibfnamefont {G.~S.}\ \bibnamefont {Barron}}, \bibinfo
		{author} {\bibfnamefont {H.~R.}\ \bibnamefont {Grimsley}}, \bibinfo {author}
		{\bibfnamefont {N.~J.}\ \bibnamefont {Mayhall}}, \bibinfo {author}
		{\bibfnamefont {E.}~\bibnamefont {Barnes}},\ and\ \bibinfo {author}
		{\bibfnamefont {S.~E.}\ \bibnamefont {Economou}},\ }\bibfield  {title}
	{\bibinfo {title} {qubit-adapt-vqe: An adaptive algorithm for constructing
			hardware-efficient ans{\"a}tze on a quantum processor},\ }\href
	{https://doi.org/https://doi.org/10.1103/PRXQuantum.2.020310} {\bibfield
		{journal} {\bibinfo  {journal} {PRX Quantum}\ }\textbf {\bibinfo {volume}
			{2}},\ \bibinfo {pages} {020310} (\bibinfo {year} {2021})}\BibitemShut
	{NoStop}%
	\bibitem [{\citenamefont {Quantum}\ \emph {et~al.}(2020)\citenamefont
		{Quantum}, \citenamefont {Collaborators}, \citenamefont {Arute},
		\citenamefont {Arya}, \citenamefont {Babbush}, \citenamefont {Bacon},
		\citenamefont {Bardin}, \citenamefont {Barends}, \citenamefont {Boixo},
		\citenamefont {Broughton}, \citenamefont {Buckley} \emph
		{et~al.}}]{google2020hartree}%
	\BibitemOpen
	\bibfield  {author} {\bibinfo {author} {\bibfnamefont {G.~A.}\ \bibnamefont
			{Quantum}}, \bibinfo {author} {\bibnamefont {Collaborators}}, \bibinfo
		{author} {\bibfnamefont {F.}~\bibnamefont {Arute}}, \bibinfo {author}
		{\bibfnamefont {K.}~\bibnamefont {Arya}}, \bibinfo {author} {\bibfnamefont
			{R.}~\bibnamefont {Babbush}}, \bibinfo {author} {\bibfnamefont
			{D.}~\bibnamefont {Bacon}}, \bibinfo {author} {\bibfnamefont {J.~C.}\
			\bibnamefont {Bardin}}, \bibinfo {author} {\bibfnamefont {R.}~\bibnamefont
			{Barends}}, \bibinfo {author} {\bibfnamefont {S.}~\bibnamefont {Boixo}},
		\bibinfo {author} {\bibfnamefont {M.}~\bibnamefont {Broughton}}, \bibinfo
		{author} {\bibfnamefont {B.~B.}\ \bibnamefont {Buckley}}, \emph {et~al.},\
	}\bibfield  {title} {\bibinfo {title} {Hartree-fock on a superconducting
			qubit quantum computer},\ }\href
	{https://doi.org/DOI:10.1126/science.abb9811} {\bibfield  {journal} {\bibinfo
			{journal} {Science}\ }\textbf {\bibinfo {volume} {369}},\ \bibinfo {pages}
		{1084} (\bibinfo {year} {2020})}\BibitemShut {NoStop}%
	\bibitem [{\citenamefont {Peruzzo}\ \emph {et~al.}(2014)\citenamefont
		{Peruzzo}, \citenamefont {McClean}, \citenamefont {Shadbolt}, \citenamefont
		{Yung}, \citenamefont {Zhou}, \citenamefont {Love}, \citenamefont
		{Aspuru-Guzik},\ and\ \citenamefont {O’brien}}]{peruzzo2014variational}%
	\BibitemOpen
	\bibfield  {author} {\bibinfo {author} {\bibfnamefont {A.}~\bibnamefont
			{Peruzzo}}, \bibinfo {author} {\bibfnamefont {J.}~\bibnamefont {McClean}},
		\bibinfo {author} {\bibfnamefont {P.}~\bibnamefont {Shadbolt}}, \bibinfo
		{author} {\bibfnamefont {M.-H.}\ \bibnamefont {Yung}}, \bibinfo {author}
		{\bibfnamefont {X.-Q.}\ \bibnamefont {Zhou}}, \bibinfo {author}
		{\bibfnamefont {P.~J.}\ \bibnamefont {Love}}, \bibinfo {author}
		{\bibfnamefont {A.}~\bibnamefont {Aspuru-Guzik}},\ and\ \bibinfo {author}
		{\bibfnamefont {J.~L.}\ \bibnamefont {O’brien}},\ }\bibfield  {title}
	{\bibinfo {title} {A variational eigenvalue solver on a photonic quantum
			processor},\ }\href {https://doi.org/https://doi.org/10.1038/ncomms5213}
	{\bibfield  {journal} {\bibinfo  {journal} {Nat. Commun.}\ }\textbf {\bibinfo
			{volume} {5}},\ \bibinfo {pages} {4213} (\bibinfo {year} {2014})}\BibitemShut
	{NoStop}%
	\bibitem [{\citenamefont {Liu}\ \emph {et~al.}(2019)\citenamefont {Liu},
		\citenamefont {Zhang}, \citenamefont {Wan},\ and\ \citenamefont
		{Wang}}]{liu2019variational}%
	\BibitemOpen
	\bibfield  {author} {\bibinfo {author} {\bibfnamefont {J.-G.}\ \bibnamefont
			{Liu}}, \bibinfo {author} {\bibfnamefont {Y.-H.}\ \bibnamefont {Zhang}},
		\bibinfo {author} {\bibfnamefont {Y.}~\bibnamefont {Wan}},\ and\ \bibinfo
		{author} {\bibfnamefont {L.}~\bibnamefont {Wang}},\ }\bibfield  {title}
	{\bibinfo {title} {Variational quantum eigensolver with fewer qubits},\
	}\href {https://doi.org/https://doi.org/10.1103/PhysRevResearch.1.023025}
	{\bibfield  {journal} {\bibinfo  {journal} {Phys. Rev. Res.}\ }\textbf
		{\bibinfo {volume} {1}},\ \bibinfo {pages} {023025} (\bibinfo {year}
		{2019})}\BibitemShut {NoStop}%
	\bibitem [{\citenamefont {Nam}\ \emph {et~al.}(2020)\citenamefont {Nam},
		\citenamefont {Chen}, \citenamefont {Pisenti}, \citenamefont {Wright},
		\citenamefont {Delaney}, \citenamefont {Maslov}, \citenamefont {Brown},
		\citenamefont {Allen}, \citenamefont {Amini}, \citenamefont {Apisdorf} \emph
		{et~al.}}]{nam2020ground}%
	\BibitemOpen
	\bibfield  {author} {\bibinfo {author} {\bibfnamefont {Y.}~\bibnamefont
			{Nam}}, \bibinfo {author} {\bibfnamefont {J.-S.}\ \bibnamefont {Chen}},
		\bibinfo {author} {\bibfnamefont {N.~C.}\ \bibnamefont {Pisenti}}, \bibinfo
		{author} {\bibfnamefont {K.}~\bibnamefont {Wright}}, \bibinfo {author}
		{\bibfnamefont {C.}~\bibnamefont {Delaney}}, \bibinfo {author} {\bibfnamefont
			{D.}~\bibnamefont {Maslov}}, \bibinfo {author} {\bibfnamefont {K.~R.}\
			\bibnamefont {Brown}}, \bibinfo {author} {\bibfnamefont {S.}~\bibnamefont
			{Allen}}, \bibinfo {author} {\bibfnamefont {J.~M.}\ \bibnamefont {Amini}},
		\bibinfo {author} {\bibfnamefont {J.}~\bibnamefont {Apisdorf}}, \emph
		{et~al.},\ }\bibfield  {title} {\bibinfo {title} {Ground-state energy
			estimation of the water molecule on a trapped-ion quantum computer},\ }\href
	{https://doi.org/https://doi.org/10.1038/s41534-020-0259-3} {\bibfield
		{journal} {\bibinfo  {journal} {npj Quantum Inf.}\ }\textbf {\bibinfo
			{volume} {6}},\ \bibinfo {pages} {33} (\bibinfo {year} {2020})}\BibitemShut
	{NoStop}%
	\bibitem [{\citenamefont {Hempel}\ \emph {et~al.}(2018)\citenamefont {Hempel},
		\citenamefont {Maier}, \citenamefont {Romero}, \citenamefont {McClean},
		\citenamefont {Monz}, \citenamefont {Shen}, \citenamefont {Jurcevic},
		\citenamefont {Lanyon}, \citenamefont {Love}, \citenamefont {Babbush} \emph
		{et~al.}}]{hempel2018quantum}%
	\BibitemOpen
	\bibfield  {author} {\bibinfo {author} {\bibfnamefont {C.}~\bibnamefont
			{Hempel}}, \bibinfo {author} {\bibfnamefont {C.}~\bibnamefont {Maier}},
		\bibinfo {author} {\bibfnamefont {J.}~\bibnamefont {Romero}}, \bibinfo
		{author} {\bibfnamefont {J.}~\bibnamefont {McClean}}, \bibinfo {author}
		{\bibfnamefont {T.}~\bibnamefont {Monz}}, \bibinfo {author} {\bibfnamefont
			{H.}~\bibnamefont {Shen}}, \bibinfo {author} {\bibfnamefont {P.}~\bibnamefont
			{Jurcevic}}, \bibinfo {author} {\bibfnamefont {B.~P.}\ \bibnamefont
			{Lanyon}}, \bibinfo {author} {\bibfnamefont {P.}~\bibnamefont {Love}},
		\bibinfo {author} {\bibfnamefont {R.}~\bibnamefont {Babbush}}, \emph
		{et~al.},\ }\bibfield  {title} {\bibinfo {title} {Quantum chemistry
			calculations on a trapped-ion quantum simulator},\ }\href
	{https://doi.org/DOI: https://doi.org/10.1103/PhysRevX.8.031022} {\bibfield
		{journal} {\bibinfo  {journal} {Phys. Rev. X}\ }\textbf {\bibinfo {volume}
			{8}},\ \bibinfo {pages} {031022} (\bibinfo {year} {2018})}\BibitemShut
	{NoStop}%
	\bibitem [{\citenamefont {Wang}\ and\ \citenamefont
		{Mazziotti}(2023)}]{wang2023electronic}%
	\BibitemOpen
	\bibfield  {author} {\bibinfo {author} {\bibfnamefont {Y.}~\bibnamefont
			{Wang}}\ and\ \bibinfo {author} {\bibfnamefont {D.~A.}\ \bibnamefont
			{Mazziotti}},\ }\bibfield  {title} {\bibinfo {title} {Electronic excited
			states from a variance-based contracted quantum eigensolver},\ }\href
	{https://doi.org/https://doi.org/10.1103/PhysRevA.108.022814} {\bibfield
		{journal} {\bibinfo  {journal} {Phys. Rev. A}\ }\textbf {\bibinfo {volume}
			{108}},\ \bibinfo {pages} {022814} (\bibinfo {year} {2023})}\BibitemShut
	{NoStop}%
	\bibitem [{\citenamefont {Ibe}\ \emph {et~al.}(2022)\citenamefont {Ibe},
		\citenamefont {Nakagawa}, \citenamefont {Earnest}, \citenamefont {Yamamoto},
		\citenamefont {Mitarai}, \citenamefont {Gao},\ and\ \citenamefont
		{Kobayashi}}]{ibe2022calculating}%
	\BibitemOpen
	\bibfield  {author} {\bibinfo {author} {\bibfnamefont {Y.}~\bibnamefont
			{Ibe}}, \bibinfo {author} {\bibfnamefont {Y.~O.}\ \bibnamefont {Nakagawa}},
		\bibinfo {author} {\bibfnamefont {N.}~\bibnamefont {Earnest}}, \bibinfo
		{author} {\bibfnamefont {T.}~\bibnamefont {Yamamoto}}, \bibinfo {author}
		{\bibfnamefont {K.}~\bibnamefont {Mitarai}}, \bibinfo {author} {\bibfnamefont
			{Q.}~\bibnamefont {Gao}},\ and\ \bibinfo {author} {\bibfnamefont
			{T.}~\bibnamefont {Kobayashi}},\ }\bibfield  {title} {\bibinfo {title}
		{Calculating transition amplitudes by variational quantum deflation},\ }\href
	{https://doi.org/10.1103/PhysRevResearch.4.013173} {\bibfield  {journal}
		{\bibinfo  {journal} {Phys. Rev. Res.}\ }\textbf {\bibinfo {volume} {4}},\
		\bibinfo {pages} {013173} (\bibinfo {year} {2022})}\BibitemShut {NoStop}%
	\bibitem [{\citenamefont {Higgott}\ \emph {et~al.}(2019)\citenamefont
		{Higgott}, \citenamefont {Wang},\ and\ \citenamefont
		{Brierley}}]{higgott2019variational}%
	\BibitemOpen
	\bibfield  {author} {\bibinfo {author} {\bibfnamefont {O.}~\bibnamefont
			{Higgott}}, \bibinfo {author} {\bibfnamefont {D.}~\bibnamefont {Wang}},\ and\
		\bibinfo {author} {\bibfnamefont {S.}~\bibnamefont {Brierley}},\ }\bibfield
	{title} {\bibinfo {title} {Variational quantum computation of excited
			states},\ }\href {https://doi.org/https://doi.org/10.22331/q-2019-07-01-156}
	{\bibfield  {journal} {\bibinfo  {journal} {Quantum}\ }\textbf {\bibinfo
			{volume} {3}},\ \bibinfo {pages} {156} (\bibinfo {year} {2019})}\BibitemShut
	{NoStop}%
	\bibitem [{\citenamefont {Jones}\ \emph {et~al.}(2019)\citenamefont {Jones},
		\citenamefont {Endo}, \citenamefont {McArdle}, \citenamefont {Yuan},\ and\
		\citenamefont {Benjamin}}]{jones2019variational}%
	\BibitemOpen
	\bibfield  {author} {\bibinfo {author} {\bibfnamefont {T.}~\bibnamefont
			{Jones}}, \bibinfo {author} {\bibfnamefont {S.}~\bibnamefont {Endo}},
		\bibinfo {author} {\bibfnamefont {S.}~\bibnamefont {McArdle}}, \bibinfo
		{author} {\bibfnamefont {X.}~\bibnamefont {Yuan}},\ and\ \bibinfo {author}
		{\bibfnamefont {S.~C.}\ \bibnamefont {Benjamin}},\ }\bibfield  {title}
	{\bibinfo {title} {Variational quantum algorithms for discovering hamiltonian
			spectra},\ }\href {https://doi.org/DOI:
		https://doi.org/10.1103/PhysRevA.99.062304} {\bibfield  {journal} {\bibinfo
			{journal} {Phys. Rev. A}\ }\textbf {\bibinfo {volume} {99}},\ \bibinfo
		{pages} {062304} (\bibinfo {year} {2019})}\BibitemShut {NoStop}%
	\bibitem [{\citenamefont {Benavides-Riveros}\ \emph {et~al.}(2024)\citenamefont
		{Benavides-Riveros}, \citenamefont {Wang}, \citenamefont {Warren},\ and\
		\citenamefont {Mazziotti}}]{benavides2024quantum}%
	\BibitemOpen
	\bibfield  {author} {\bibinfo {author} {\bibfnamefont {C.~L.}\ \bibnamefont
			{Benavides-Riveros}}, \bibinfo {author} {\bibfnamefont {Y.}~\bibnamefont
			{Wang}}, \bibinfo {author} {\bibfnamefont {S.}~\bibnamefont {Warren}},\ and\
		\bibinfo {author} {\bibfnamefont {D.~A.}\ \bibnamefont {Mazziotti}},\
	}\bibfield  {title} {\bibinfo {title} {Quantum simulation of excited states
			from parallel contracted quantum eigensolvers},\ }\href
	{https://doi.org/10.1088/1367-2630/ad2d1d} {\bibfield  {journal} {\bibinfo
			{journal} {New J. Phys.}\ }\textbf {\bibinfo {volume} {26}},\ \bibinfo
		{pages} {033020} (\bibinfo {year} {2024})}\BibitemShut {NoStop}%
	\bibitem [{\citenamefont {Smart}\ \emph {et~al.}(2024)\citenamefont {Smart},
		\citenamefont {Welakuh},\ and\ \citenamefont {Narang}}]{smart2024many}%
	\BibitemOpen
	\bibfield  {author} {\bibinfo {author} {\bibfnamefont {S.~E.}\ \bibnamefont
			{Smart}}, \bibinfo {author} {\bibfnamefont {D.~M.}\ \bibnamefont {Welakuh}},\
		and\ \bibinfo {author} {\bibfnamefont {P.}~\bibnamefont {Narang}},\
	}\bibfield  {title} {\bibinfo {title} {Many-body excited states with a
			contracted quantum eigensolver},\ }\href
	{https://doi.org/https://doi.org/10.1021/acs.jctc.4c00030} {\bibfield
		{journal} {\bibinfo  {journal} {J. Chem. Theory Comput.}\ }\textbf {\bibinfo
			{volume} {20}},\ \bibinfo {pages} {3580} (\bibinfo {year}
		{2024})}\BibitemShut {NoStop}%
	\bibitem [{\citenamefont {Jouzdani}\ \emph {et~al.}(2021)\citenamefont
		{Jouzdani}, \citenamefont {Bringuier},\ and\ \citenamefont
		{Kostuk}}]{jouzdani2021method}%
	\BibitemOpen
	\bibfield  {author} {\bibinfo {author} {\bibfnamefont {P.}~\bibnamefont
			{Jouzdani}}, \bibinfo {author} {\bibfnamefont {S.}~\bibnamefont
			{Bringuier}},\ and\ \bibinfo {author} {\bibfnamefont {M.}~\bibnamefont
			{Kostuk}},\ }\bibfield  {title} {\bibinfo {title} {A method of determining
			molecular excited-states using quantum computation},\ }\href
	{https://doi.org/https://doi.org/10.1557/s43580-021-00111-3} {\bibfield
		{journal} {\bibinfo  {journal} {MRS Adv.}\ }\textbf {\bibinfo {volume} {6}},\
		\bibinfo {pages} {558} (\bibinfo {year} {2021})}\BibitemShut {NoStop}%
	\bibitem [{\citenamefont {Nakanishi}\ \emph {et~al.}(2019)\citenamefont
		{Nakanishi}, \citenamefont {Mitarai},\ and\ \citenamefont
		{Fujii}}]{nakanishi2019subspace}%
	\BibitemOpen
	\bibfield  {author} {\bibinfo {author} {\bibfnamefont {K.~M.}\ \bibnamefont
			{Nakanishi}}, \bibinfo {author} {\bibfnamefont {K.}~\bibnamefont {Mitarai}},\
		and\ \bibinfo {author} {\bibfnamefont {K.}~\bibnamefont {Fujii}},\ }\bibfield
	{title} {\bibinfo {title} {Subspace-search variational quantum eigensolver
			for excited states},\ }\href
	{https://doi.org/https://doi.org/10.1103/PhysRevResearch.1.033062} {\bibfield
		{journal} {\bibinfo  {journal} {Phys. Rev. Res.}\ }\textbf {\bibinfo
			{volume} {1}},\ \bibinfo {pages} {033062} (\bibinfo {year}
		{2019})}\BibitemShut {NoStop}%
	\bibitem [{\citenamefont {Han}\ \emph {et~al.}(2024)\citenamefont {Han},
		\citenamefont {Lyu}, \citenamefont {Zhou}, \citenamefont {Yuan},
		\citenamefont {Chu}, \citenamefont {Nuerbolati}, \citenamefont {Jia},
		\citenamefont {Nie}, \citenamefont {Wei}, \citenamefont {Yang} \emph
		{et~al.}}]{han2024multilevel}%
	\BibitemOpen
	\bibfield  {author} {\bibinfo {author} {\bibfnamefont {Z.}~\bibnamefont
			{Han}}, \bibinfo {author} {\bibfnamefont {C.}~\bibnamefont {Lyu}}, \bibinfo
		{author} {\bibfnamefont {Y.}~\bibnamefont {Zhou}}, \bibinfo {author}
		{\bibfnamefont {J.}~\bibnamefont {Yuan}}, \bibinfo {author} {\bibfnamefont
			{J.}~\bibnamefont {Chu}}, \bibinfo {author} {\bibfnamefont {W.}~\bibnamefont
			{Nuerbolati}}, \bibinfo {author} {\bibfnamefont {H.}~\bibnamefont {Jia}},
		\bibinfo {author} {\bibfnamefont {L.}~\bibnamefont {Nie}}, \bibinfo {author}
		{\bibfnamefont {W.}~\bibnamefont {Wei}}, \bibinfo {author} {\bibfnamefont
			{Z.}~\bibnamefont {Yang}}, \emph {et~al.},\ }\bibfield  {title} {\bibinfo
		{title} {Multilevel variational spectroscopy using a programmable quantum
			simulator},\ }\href
	{https://doi.org/https://doi.org/10.1103/PhysRevResearch.6.013015} {\bibfield
		{journal} {\bibinfo  {journal} {Phys. Rev. Res.}\ }\textbf {\bibinfo
			{volume} {6}},\ \bibinfo {pages} {013015} (\bibinfo {year}
		{2024})}\BibitemShut {NoStop}%
	\bibitem [{\citenamefont {Yalouz}\ \emph {et~al.}(2021)\citenamefont {Yalouz},
		\citenamefont {Senjean}, \citenamefont {G{\"u}nther}, \citenamefont {Buda},
		\citenamefont {O’Brien},\ and\ \citenamefont {Visscher}}]{yalouz2021state}%
	\BibitemOpen
	\bibfield  {author} {\bibinfo {author} {\bibfnamefont {S.}~\bibnamefont
			{Yalouz}}, \bibinfo {author} {\bibfnamefont {B.}~\bibnamefont {Senjean}},
		\bibinfo {author} {\bibfnamefont {J.}~\bibnamefont {G{\"u}nther}}, \bibinfo
		{author} {\bibfnamefont {F.}~\bibnamefont {Buda}}, \bibinfo {author}
		{\bibfnamefont {T.~E.}\ \bibnamefont {O’Brien}},\ and\ \bibinfo {author}
		{\bibfnamefont {L.}~\bibnamefont {Visscher}},\ }\bibfield  {title} {\bibinfo
		{title} {A state-averaged orbital-optimized hybrid quantum--classical
			algorithm for a democratic description of ground and excited states},\ }\href
	{https://doi.org/10.1088/2058-9565/abd334} {\bibfield  {journal} {\bibinfo
			{journal} {Quantum. Sci. Technol.}\ }\textbf {\bibinfo {volume} {6}},\
		\bibinfo {pages} {024004} (\bibinfo {year} {2021})}\BibitemShut {NoStop}%
	\bibitem [{\citenamefont {Parrish}\ \emph {et~al.}(2019)\citenamefont
		{Parrish}, \citenamefont {Hohenstein}, \citenamefont {McMahon},\ and\
		\citenamefont {Mart{\'\i}nez}}]{parrish2019quantum}%
	\BibitemOpen
	\bibfield  {author} {\bibinfo {author} {\bibfnamefont {R.~M.}\ \bibnamefont
			{Parrish}}, \bibinfo {author} {\bibfnamefont {E.~G.}\ \bibnamefont
			{Hohenstein}}, \bibinfo {author} {\bibfnamefont {P.~L.}\ \bibnamefont
			{McMahon}},\ and\ \bibinfo {author} {\bibfnamefont {T.~J.}\ \bibnamefont
			{Mart{\'\i}nez}},\ }\bibfield  {title} {\bibinfo {title} {Quantum computation
			of electronic transitions using a variational quantum eigensolver},\ }\href
	{https://doi.org/DOI: 10.1103/PhysRevLett.122.230401} {\bibfield  {journal}
		{\bibinfo  {journal} {Phys. Rev. Lett.}\ }\textbf {\bibinfo {volume} {122}},\
		\bibinfo {pages} {230401} (\bibinfo {year} {2019})}\BibitemShut {NoStop}%
	\bibitem [{\citenamefont {Dutta}\ \emph {et~al.}(2025)\citenamefont {Dutta},
		\citenamefont {Cianci}, \citenamefont {Soudackov}, \citenamefont {Wang},
		\citenamefont {Xu}, \citenamefont {Mazziotti}, \citenamefont {Santos},\ and\
		\citenamefont {Batista}}]{dutta2025qumode}%
	\BibitemOpen
	\bibfield  {author} {\bibinfo {author} {\bibfnamefont {R.}~\bibnamefont
			{Dutta}}, \bibinfo {author} {\bibfnamefont {C.}~\bibnamefont {Cianci}},
		\bibinfo {author} {\bibfnamefont {A.~V.}\ \bibnamefont {Soudackov}}, \bibinfo
		{author} {\bibfnamefont {Y.}~\bibnamefont {Wang}}, \bibinfo {author}
		{\bibfnamefont {C.}~\bibnamefont {Xu}}, \bibinfo {author} {\bibfnamefont
			{D.~A.}\ \bibnamefont {Mazziotti}}, \bibinfo {author} {\bibfnamefont {L.~F.}\
			\bibnamefont {Santos}},\ and\ \bibinfo {author} {\bibfnamefont {V.~S.}\
			\bibnamefont {Batista}},\ }\bibfield  {title} {\bibinfo {title} {Qumode-based
			variational quantum eigensolver for molecular excited states},\ }\href@noop
	{} {\bibfield  {journal} {\bibinfo  {journal} {preprint arXiv:2509.04727}\ }
		(\bibinfo {year} {2025})}\BibitemShut {NoStop}%
	\bibitem [{\citenamefont {Guo}\ \emph {et~al.}(2024)\citenamefont {Guo},
		\citenamefont {Angelides}, \citenamefont {Jansen},\ and\ \citenamefont
		{K{\"u}hn}}]{guo2024concurrent}%
	\BibitemOpen
	\bibfield  {author} {\bibinfo {author} {\bibfnamefont {Y.}~\bibnamefont
			{Guo}}, \bibinfo {author} {\bibfnamefont {T.}~\bibnamefont {Angelides}},
		\bibinfo {author} {\bibfnamefont {K.}~\bibnamefont {Jansen}},\ and\ \bibinfo
		{author} {\bibfnamefont {S.}~\bibnamefont {K{\"u}hn}},\ }\bibfield  {title}
	{\bibinfo {title} {Concurrent vqe for simulating excited states of the
			schwinger model},\ }\href@noop {} {\bibfield  {journal} {\bibinfo  {journal}
			{preprint arXiv:2407.15629}\ } (\bibinfo {year} {2024})}\BibitemShut
	{NoStop}%
	\bibitem [{\citenamefont {Xu}\ \emph {et~al.}(2023)\citenamefont {Xu},
		\citenamefont {Guo}, \citenamefont {Li}, \citenamefont {Wang}, \citenamefont
		{Fan}, \citenamefont {Zhou}, \citenamefont {Liao},\ and\ \citenamefont
		{Xiang}}]{xu2023concurrent}%
	\BibitemOpen
	\bibfield  {author} {\bibinfo {author} {\bibfnamefont {G.}~\bibnamefont
			{Xu}}, \bibinfo {author} {\bibfnamefont {Y.}~\bibnamefont {Guo}}, \bibinfo
		{author} {\bibfnamefont {X.}~\bibnamefont {Li}}, \bibinfo {author}
		{\bibfnamefont {K.}~\bibnamefont {Wang}}, \bibinfo {author} {\bibfnamefont
			{Z.}~\bibnamefont {Fan}}, \bibinfo {author} {\bibfnamefont {Z.}~\bibnamefont
			{Zhou}}, \bibinfo {author} {\bibfnamefont {H.}~\bibnamefont {Liao}},\ and\
		\bibinfo {author} {\bibfnamefont {T.}~\bibnamefont {Xiang}},\ }\bibfield
	{title} {\bibinfo {title} {Concurrent quantum eigensolver for multiple
			low-energy eigenstates},\ }\href
	{https://doi.org/https://doi.org/10.1103/PhysRevA.107.052423} {\bibfield
		{journal} {\bibinfo  {journal} {Phys. Rev. A}\ }\textbf {\bibinfo {volume}
			{107}},\ \bibinfo {pages} {052423} (\bibinfo {year} {2023})}\BibitemShut
	{NoStop}%
	\bibitem [{\citenamefont {BAQIS}(2024)}]{baqisQuafuSuperconducting}%
	\BibitemOpen
	\bibfield  {author} {\bibinfo {author} {\bibnamefont {BAQIS}},\ }\href@noop
	{} {\bibinfo {title} {Quafu superconducting quantum computing}} (\bibinfo
	{year} {2024}),\ \bibinfo {note}
	{\url{https://quafu-sqc.baqis.ac.cn}}\BibitemShut {NoStop}%
	\bibitem [{\citenamefont {Spall}(1992)}]{spall1992multivariate}%
	\BibitemOpen
	\bibfield  {author} {\bibinfo {author} {\bibfnamefont {J.~C.}\ \bibnamefont
			{Spall}},\ }\bibfield  {title} {\bibinfo {title} {Multivariate stochastic
			approximation using a simultaneous perturbation gradient approximation},\
	}\href {https://doi.org/10.1109/9.119632} {\bibfield  {journal} {\bibinfo
			{journal} {IEEE Trans. Autom.}\ }\textbf {\bibinfo {volume} {37}},\ \bibinfo
		{pages} {332} (\bibinfo {year} {1992})}\BibitemShut {NoStop}%
	\bibitem [{\citenamefont {Hong}\ \emph {et~al.}(2024)\citenamefont {Hong},
		\citenamefont {Colmenarez}, \citenamefont {Ding}, \citenamefont
		{Benavides-Riveros},\ and\ \citenamefont {Schilling}}]{hong2024Refining}%
	\BibitemOpen
	\bibfield  {author} {\bibinfo {author} {\bibfnamefont {C.-L.}\ \bibnamefont
			{Hong}}, \bibinfo {author} {\bibfnamefont {L.}~\bibnamefont {Colmenarez}},
		\bibinfo {author} {\bibfnamefont {L.}~\bibnamefont {Ding}}, \bibinfo {author}
		{\bibfnamefont {C.~L.}\ \bibnamefont {Benavides-Riveros}},\ and\ \bibinfo
		{author} {\bibfnamefont {C.}~\bibnamefont {Schilling}},\ }\bibfield  {title}
	{\bibinfo {title} {Refining the weighted subspace-search variational quantum
			eigensolver: compression of ansatze into a single pure state and optimization
			of weights},\ }\href@noop {} {\bibfield  {journal} {\bibinfo  {journal}
			{preprint arXiv: 2306.11844}\ } (\bibinfo {year} {2024})}\BibitemShut
	{NoStop}%
	\bibitem [{\citenamefont {McClean}\ \emph {et~al.}(2020)\citenamefont
		{McClean}, \citenamefont {Rubin}, \citenamefont {Sung}, \citenamefont
		{Kivlichan}, \citenamefont {Bonet-Monroig}, \citenamefont {Cao},
		\citenamefont {Dai}, \citenamefont {Fried}, \citenamefont {Gidney},
		\citenamefont {Gimby} \emph {et~al.}}]{mcclean2020openfermion}%
	\BibitemOpen
	\bibfield  {author} {\bibinfo {author} {\bibfnamefont {J.~R.}\ \bibnamefont
			{McClean}}, \bibinfo {author} {\bibfnamefont {N.~C.}\ \bibnamefont {Rubin}},
		\bibinfo {author} {\bibfnamefont {K.~J.}\ \bibnamefont {Sung}}, \bibinfo
		{author} {\bibfnamefont {I.~D.}\ \bibnamefont {Kivlichan}}, \bibinfo {author}
		{\bibfnamefont {X.}~\bibnamefont {Bonet-Monroig}}, \bibinfo {author}
		{\bibfnamefont {Y.}~\bibnamefont {Cao}}, \bibinfo {author} {\bibfnamefont
			{C.}~\bibnamefont {Dai}}, \bibinfo {author} {\bibfnamefont {E.~S.}\
			\bibnamefont {Fried}}, \bibinfo {author} {\bibfnamefont {C.}~\bibnamefont
			{Gidney}}, \bibinfo {author} {\bibfnamefont {B.}~\bibnamefont {Gimby}}, \emph
		{et~al.},\ }\bibfield  {title} {\bibinfo {title} {Openfermion: the electronic
			structure package for quantum computers},\ }\href
	{https://doi.org/10.1088/2058-9565/ab8ebc} {\bibfield  {journal} {\bibinfo
			{journal} {Quantum Sci. and Technol.}\ }\textbf {\bibinfo {volume} {5}},\
		\bibinfo {pages} {034014} (\bibinfo {year} {2020})}\BibitemShut {NoStop}%
	\bibitem [{\citenamefont {Sun}\ \emph {et~al.}(2018)\citenamefont {Sun},
		\citenamefont {Berkelbach}, \citenamefont {Blunt}, \citenamefont {Booth},
		\citenamefont {Guo}, \citenamefont {Li}, \citenamefont {Liu}, \citenamefont
		{McClain}, \citenamefont {Sayfutyarova}, \citenamefont {Sharma} \emph
		{et~al.}}]{sun2018pyscf}%
	\BibitemOpen
	\bibfield  {author} {\bibinfo {author} {\bibfnamefont {Q.}~\bibnamefont
			{Sun}}, \bibinfo {author} {\bibfnamefont {T.~C.}\ \bibnamefont {Berkelbach}},
		\bibinfo {author} {\bibfnamefont {N.~S.}\ \bibnamefont {Blunt}}, \bibinfo
		{author} {\bibfnamefont {G.~H.}\ \bibnamefont {Booth}}, \bibinfo {author}
		{\bibfnamefont {S.}~\bibnamefont {Guo}}, \bibinfo {author} {\bibfnamefont
			{Z.}~\bibnamefont {Li}}, \bibinfo {author} {\bibfnamefont {J.}~\bibnamefont
			{Liu}}, \bibinfo {author} {\bibfnamefont {J.~D.}\ \bibnamefont {McClain}},
		\bibinfo {author} {\bibfnamefont {E.~R.}\ \bibnamefont {Sayfutyarova}},
		\bibinfo {author} {\bibfnamefont {S.}~\bibnamefont {Sharma}}, \emph
		{et~al.},\ }\bibfield  {title} {\bibinfo {title} {Pyscf: the python-based
			simulations of chemistry framework},\ }\href
	{https://doi.org/10.1002/wcms.1340} {\bibfield  {journal} {\bibinfo
			{journal} {Wires. Comput. Mol. Sci.}\ }\textbf {\bibinfo {volume} {8}},\
		\bibinfo {pages} {e1340} (\bibinfo {year} {2018})}\BibitemShut {NoStop}%
	\bibitem [{\citenamefont {Pfeuty}(1970)}]{pfeuty1970one}%
	\BibitemOpen
	\bibfield  {author} {\bibinfo {author} {\bibfnamefont {P.}~\bibnamefont
			{Pfeuty}},\ }\bibfield  {title} {\bibinfo {title} {The one-dimensional ising
			model with a transverse field},\ }\href
	{https://doi.org/https://doi.org/10.1016/0003-4916(70)90270-8} {\bibfield
		{journal} {\bibinfo  {journal} {Ann. Phys.}\ }\textbf {\bibinfo {volume}
			{57}},\ \bibinfo {pages} {79} (\bibinfo {year} {1970})}\BibitemShut {NoStop}%
	\bibitem [{\citenamefont {Heyl}\ \emph {et~al.}(2013)\citenamefont {Heyl},
		\citenamefont {Polkovnikov},\ and\ \citenamefont
		{Kehrein}}]{heyl2013dynamical}%
	\BibitemOpen
	\bibfield  {author} {\bibinfo {author} {\bibfnamefont {M.}~\bibnamefont
			{Heyl}}, \bibinfo {author} {\bibfnamefont {A.}~\bibnamefont {Polkovnikov}},\
		and\ \bibinfo {author} {\bibfnamefont {S.}~\bibnamefont {Kehrein}},\
	}\bibfield  {title} {\bibinfo {title} {Dynamical quantum phase transitions in
			the transverse-field ising model},\ }\href
	{https://doi.org/https://doi.org/10.1103/PhysRevLett.110.135704} {\bibfield
		{journal} {\bibinfo  {journal} {Phys. Rev. Lett.}\ }\textbf {\bibinfo
			{volume} {110}},\ \bibinfo {pages} {135704} (\bibinfo {year}
		{2013})}\BibitemShut {NoStop}%
	\bibitem [{\citenamefont {Mondaini}\ \emph {et~al.}(2016)\citenamefont
		{Mondaini}, \citenamefont {Fratus}, \citenamefont {Srednicki},\ and\
		\citenamefont {Rigol}}]{mondaini2016eigenstate}%
	\BibitemOpen
	\bibfield  {author} {\bibinfo {author} {\bibfnamefont {R.}~\bibnamefont
			{Mondaini}}, \bibinfo {author} {\bibfnamefont {K.~R.}\ \bibnamefont
			{Fratus}}, \bibinfo {author} {\bibfnamefont {M.}~\bibnamefont {Srednicki}},\
		and\ \bibinfo {author} {\bibfnamefont {M.}~\bibnamefont {Rigol}},\ }\bibfield
	{title} {\bibinfo {title} {Eigenstate thermalization in the two-dimensional
			transverse field ising model},\ }\href {https://doi.org/DOI:
		https://doi.org/10.1103/PhysRevE.93.032104} {\bibfield  {journal} {\bibinfo
			{journal} {Phys. Rev. E}\ }\textbf {\bibinfo {volume} {93}},\ \bibinfo
		{pages} {032104} (\bibinfo {year} {2016})}\BibitemShut {NoStop}%
	\bibitem [{\citenamefont {Schmitt}\ \emph {et~al.}(2022)\citenamefont
		{Schmitt}, \citenamefont {Rams}, \citenamefont {Dziarmaga}, \citenamefont
		{Heyl},\ and\ \citenamefont {Zurek}}]{schmitt2022quantum}%
	\BibitemOpen
	\bibfield  {author} {\bibinfo {author} {\bibfnamefont {M.}~\bibnamefont
			{Schmitt}}, \bibinfo {author} {\bibfnamefont {M.~M.}\ \bibnamefont {Rams}},
		\bibinfo {author} {\bibfnamefont {J.}~\bibnamefont {Dziarmaga}}, \bibinfo
		{author} {\bibfnamefont {M.}~\bibnamefont {Heyl}},\ and\ \bibinfo {author}
		{\bibfnamefont {W.~H.}\ \bibnamefont {Zurek}},\ }\bibfield  {title} {\bibinfo
		{title} {Quantum phase transition dynamics in the two-dimensional
			transverse-field ising model},\ }\href
	{https://doi.org/DOI:10.1126/sciadv.abl6850} {\bibfield  {journal} {\bibinfo
			{journal} {Sci. Adv.}\ }\textbf {\bibinfo {volume} {8}},\ \bibinfo {pages}
		{eabl6850} (\bibinfo {year} {2022})}\BibitemShut {NoStop}%
	\bibitem [{\citenamefont {Li}\ \emph {et~al.}(2023)\citenamefont {Li},
		\citenamefont {Wu}, \citenamefont {Mei}, \citenamefont {Yao}, \citenamefont
		{Lian}, \citenamefont {Cai}, \citenamefont {Wang}, \citenamefont {Qi},
		\citenamefont {Yao}, \citenamefont {He} \emph {et~al.}}]{li2023probing}%
	\BibitemOpen
	\bibfield  {author} {\bibinfo {author} {\bibfnamefont {B.-W.}\ \bibnamefont
			{Li}}, \bibinfo {author} {\bibfnamefont {Y.-K.}\ \bibnamefont {Wu}}, \bibinfo
		{author} {\bibfnamefont {Q.-X.}\ \bibnamefont {Mei}}, \bibinfo {author}
		{\bibfnamefont {R.}~\bibnamefont {Yao}}, \bibinfo {author} {\bibfnamefont
			{W.-Q.}\ \bibnamefont {Lian}}, \bibinfo {author} {\bibfnamefont {M.-L.}\
			\bibnamefont {Cai}}, \bibinfo {author} {\bibfnamefont {Y.}~\bibnamefont
			{Wang}}, \bibinfo {author} {\bibfnamefont {B.-X.}\ \bibnamefont {Qi}},
		\bibinfo {author} {\bibfnamefont {L.}~\bibnamefont {Yao}}, \bibinfo {author}
		{\bibfnamefont {L.}~\bibnamefont {He}}, \emph {et~al.},\ }\bibfield  {title}
	{\bibinfo {title} {Probing critical behavior of long-range transverse-field
			ising model through quantum kibble-zurek mechanism},\ }\href
	{https://doi.org/https://doi.org/10.1103/PRXQuantum.4.010302} {\bibfield
		{journal} {\bibinfo  {journal} {PRX Quantum}\ }\textbf {\bibinfo {volume}
			{4}},\ \bibinfo {pages} {010302} (\bibinfo {year} {2023})}\BibitemShut
	{NoStop}%
	\bibitem [{\citenamefont {Cai}\ \emph {et~al.}(2023)\citenamefont {Cai},
		\citenamefont {Babbush}, \citenamefont {Benjamin}, \citenamefont {Endo},
		\citenamefont {Huggins}, \citenamefont {Li}, \citenamefont {McClean},\ and\
		\citenamefont {O’Brien}}]{cai2023quantum}%
	\BibitemOpen
	\bibfield  {author} {\bibinfo {author} {\bibfnamefont {Z.}~\bibnamefont
			{Cai}}, \bibinfo {author} {\bibfnamefont {R.}~\bibnamefont {Babbush}},
		\bibinfo {author} {\bibfnamefont {S.~C.}\ \bibnamefont {Benjamin}}, \bibinfo
		{author} {\bibfnamefont {S.}~\bibnamefont {Endo}}, \bibinfo {author}
		{\bibfnamefont {W.~J.}\ \bibnamefont {Huggins}}, \bibinfo {author}
		{\bibfnamefont {Y.}~\bibnamefont {Li}}, \bibinfo {author} {\bibfnamefont
			{J.~R.}\ \bibnamefont {McClean}},\ and\ \bibinfo {author} {\bibfnamefont
			{T.~E.}\ \bibnamefont {O’Brien}},\ }\bibfield  {title} {\bibinfo {title}
		{Quantum error mitigation},\ }\href
	{https://doi.org/https://doi.org/10.1103/RevModPhys.95.045005} {\bibfield
		{journal} {\bibinfo  {journal} {Rev. Mod. Phys.}\ }\textbf {\bibinfo {volume}
			{95}},\ \bibinfo {pages} {045005} (\bibinfo {year} {2023})}\BibitemShut
	{NoStop}%
	\bibitem [{\citenamefont {Bonet-Monroig}\ \emph {et~al.}(2018)\citenamefont
		{Bonet-Monroig}, \citenamefont {Sagastizabal}, \citenamefont {Singh},\ and\
		\citenamefont {O'Brien}}]{bonet2018low}%
	\BibitemOpen
	\bibfield  {author} {\bibinfo {author} {\bibfnamefont {X.}~\bibnamefont
			{Bonet-Monroig}}, \bibinfo {author} {\bibfnamefont {R.}~\bibnamefont
			{Sagastizabal}}, \bibinfo {author} {\bibfnamefont {M.}~\bibnamefont
			{Singh}},\ and\ \bibinfo {author} {\bibfnamefont {T.}~\bibnamefont
			{O'Brien}},\ }\bibfield  {title} {\bibinfo {title} {Low-cost error mitigation
			by symmetry verification},\ }\href {https://doi.org/DOI:
		https://doi.org/10.1103/PhysRevA.98.062339} {\bibfield  {journal} {\bibinfo
			{journal} {Phys. Rev. A}\ }\textbf {\bibinfo {volume} {98}},\ \bibinfo
		{pages} {062339} (\bibinfo {year} {2018})}\BibitemShut {NoStop}%
	\bibitem [{\citenamefont {Islam}\ \emph {et~al.}(2011)\citenamefont {Islam},
		\citenamefont {Edwards}, \citenamefont {Kim}, \citenamefont {Korenblit},
		\citenamefont {Noh}, \citenamefont {Carmichael}, \citenamefont {Lin},
		\citenamefont {Duan}, \citenamefont {Joseph~Wang}, \citenamefont {Freericks}
		\emph {et~al.}}]{islam2011onset}%
	\BibitemOpen
	\bibfield  {author} {\bibinfo {author} {\bibfnamefont {R.}~\bibnamefont
			{Islam}}, \bibinfo {author} {\bibfnamefont {E.}~\bibnamefont {Edwards}},
		\bibinfo {author} {\bibfnamefont {K.}~\bibnamefont {Kim}}, \bibinfo {author}
		{\bibfnamefont {S.}~\bibnamefont {Korenblit}}, \bibinfo {author}
		{\bibfnamefont {C.}~\bibnamefont {Noh}}, \bibinfo {author} {\bibfnamefont
			{H.}~\bibnamefont {Carmichael}}, \bibinfo {author} {\bibfnamefont {G.-D.}\
			\bibnamefont {Lin}}, \bibinfo {author} {\bibfnamefont {L.-M.}\ \bibnamefont
			{Duan}}, \bibinfo {author} {\bibfnamefont {C.-C.}\ \bibnamefont
			{Joseph~Wang}}, \bibinfo {author} {\bibfnamefont {J.}~\bibnamefont
			{Freericks}}, \emph {et~al.},\ }\bibfield  {title} {\bibinfo {title} {Onset
			of a quantum phase transition with a trapped ion quantum simulator},\ }\href
	{https://doi.org/https://doi.org/10.1038/ncomms1374} {\bibfield  {journal}
		{\bibinfo  {journal} {Nat. Commun.}\ }\textbf {\bibinfo {volume} {2}},\
		\bibinfo {pages} {377} (\bibinfo {year} {2011})}\BibitemShut {NoStop}%
	\bibitem [{\citenamefont {Binder}(1981)}]{binder1981critical}%
	\BibitemOpen
	\bibfield  {author} {\bibinfo {author} {\bibfnamefont {K.}~\bibnamefont
			{Binder}},\ }\bibfield  {title} {\bibinfo {title} {Critical properties from
			monte carlo coarse graining and renormalization},\ }\href
	{https://journals.aps.org/prl/abstract/10.1103/PhysRevLett.47.693} {\bibfield
		{journal} {\bibinfo  {journal} {Phys. Rev. Lett.}\ }\textbf {\bibinfo
			{volume} {47}},\ \bibinfo {pages} {693} (\bibinfo {year} {1981})}\BibitemShut
	{NoStop}%
	\bibitem [{\citenamefont {Fisher}\ and\ \citenamefont
		{Barber}(1972)}]{fisher1972scaling}%
	\BibitemOpen
	\bibfield  {author} {\bibinfo {author} {\bibfnamefont {M.~E.}\ \bibnamefont
			{Fisher}}\ and\ \bibinfo {author} {\bibfnamefont {M.~N.}\ \bibnamefont
			{Barber}},\ }\bibfield  {title} {\bibinfo {title} {Scaling theory for
			finite-size effects in the critical region},\ }\href
	{https://journals.aps.org/prl/abstract/10.1103/PhysRevLett.28.1516}
	{\bibfield  {journal} {\bibinfo  {journal} {Phys. Rev. Lett.}\ }\textbf
		{\bibinfo {volume} {28}},\ \bibinfo {pages} {1516} (\bibinfo {year}
		{1972})}\BibitemShut {NoStop}%
	\bibitem [{\citenamefont {Cross}(2018)}]{qiskit}%
	\BibitemOpen
	\bibfield  {author} {\bibinfo {author} {\bibfnamefont {A.}~\bibnamefont
			{Cross}},\ }\bibfield  {title} {\bibinfo {title} {The {IBM Q} experience and
			{Qiskit} open-source quantum computing software},\ }\href
	{https://meetings.aps.org/Meeting/MAR18/Event/323385} {\bibfield  {journal}
		{\bibinfo  {journal} {Bulletin of the American Physical Society}\ }\textbf
		{\bibinfo {volume} {63}} (\bibinfo {year} {2018})}\BibitemShut {NoStop}%
	\bibitem [{\citenamefont {Stilck~Fran{\c{c}}a}\ and\ \citenamefont
		{Garcia-Patron}(2021)}]{stilck2021limitations}%
	\BibitemOpen
	\bibfield  {author} {\bibinfo {author} {\bibfnamefont {D.}~\bibnamefont
			{Stilck~Fran{\c{c}}a}}\ and\ \bibinfo {author} {\bibfnamefont
			{R.}~\bibnamefont {Garcia-Patron}},\ }\bibfield  {title} {\bibinfo {title}
		{Limitations of optimization algorithms on noisy quantum devices},\ }\href
	{https://www.nature.com/articles/s41567-021-01356-3} {\bibfield  {journal}
		{\bibinfo  {journal} {Nat. Phys.}\ }\textbf {\bibinfo {volume} {17}},\
		\bibinfo {pages} {1221} (\bibinfo {year} {2021})}\BibitemShut {NoStop}%
	\bibitem [{\citenamefont {Sharma}\ \emph {et~al.}(2022)\citenamefont {Sharma},
		\citenamefont {Cerezo}, \citenamefont {Cincio},\ and\ \citenamefont
		{Coles}}]{sharma2022trainability}%
	\BibitemOpen
	\bibfield  {author} {\bibinfo {author} {\bibfnamefont {K.}~\bibnamefont
			{Sharma}}, \bibinfo {author} {\bibfnamefont {M.}~\bibnamefont {Cerezo}},
		\bibinfo {author} {\bibfnamefont {L.}~\bibnamefont {Cincio}},\ and\ \bibinfo
		{author} {\bibfnamefont {P.~J.}\ \bibnamefont {Coles}},\ }\bibfield  {title}
	{\bibinfo {title} {Trainability of dissipative perceptron-based quantum
			neural networks},\ }\href
	{https://journals.aps.org/prl/abstract/10.1103/PhysRevLett.128.180505}
	{\bibfield  {journal} {\bibinfo  {journal} {Phys. Rev. Lett.}\ }\textbf
		{\bibinfo {volume} {128}},\ \bibinfo {pages} {180505} (\bibinfo {year}
		{2022})}\BibitemShut {NoStop}%
	\bibitem [{\citenamefont {Wang}\ \emph {et~al.}(2021)\citenamefont {Wang},
		\citenamefont {Fontana}, \citenamefont {Cerezo}, \citenamefont {Sharma},
		\citenamefont {Sone}, \citenamefont {Cincio},\ and\ \citenamefont
		{Coles}}]{wang2021noise}%
	\BibitemOpen
	\bibfield  {author} {\bibinfo {author} {\bibfnamefont {S.}~\bibnamefont
			{Wang}}, \bibinfo {author} {\bibfnamefont {E.}~\bibnamefont {Fontana}},
		\bibinfo {author} {\bibfnamefont {M.}~\bibnamefont {Cerezo}}, \bibinfo
		{author} {\bibfnamefont {K.}~\bibnamefont {Sharma}}, \bibinfo {author}
		{\bibfnamefont {A.}~\bibnamefont {Sone}}, \bibinfo {author} {\bibfnamefont
			{L.}~\bibnamefont {Cincio}},\ and\ \bibinfo {author} {\bibfnamefont {P.~J.}\
			\bibnamefont {Coles}},\ }\bibfield  {title} {\bibinfo {title} {Noise-induced
			barren plateaus in variational quantum algorithms},\ }\href {https://www.Nat.
		Commun..com/articles/s41467-021-27045-6} {\bibfield  {journal} {\bibinfo
			{journal} {Nat. Commun.}\ }\textbf {\bibinfo {volume} {12}},\ \bibinfo
		{pages} {6961} (\bibinfo {year} {2021})}\BibitemShut {NoStop}%
	\bibitem [{\citenamefont {McClean}\ \emph {et~al.}(2018)\citenamefont
		{McClean}, \citenamefont {Boixo}, \citenamefont {Smelyanskiy}, \citenamefont
		{Babbush},\ and\ \citenamefont {Neven}}]{mcclean2018barren}%
	\BibitemOpen
	\bibfield  {author} {\bibinfo {author} {\bibfnamefont {J.~R.}\ \bibnamefont
			{McClean}}, \bibinfo {author} {\bibfnamefont {S.}~\bibnamefont {Boixo}},
		\bibinfo {author} {\bibfnamefont {V.~N.}\ \bibnamefont {Smelyanskiy}},
		\bibinfo {author} {\bibfnamefont {R.}~\bibnamefont {Babbush}},\ and\ \bibinfo
		{author} {\bibfnamefont {H.}~\bibnamefont {Neven}},\ }\bibfield  {title}
	{\bibinfo {title} {Barren plateaus in quantum neural network training
			landscapes},\ }\href {https://www.Nat.
		Commun..com/articles/s41467-018-07090-4} {\bibfield  {journal} {\bibinfo
			{journal} {Nat. Commun.}\ }\textbf {\bibinfo {volume} {9}},\ \bibinfo {pages}
		{4812} (\bibinfo {year} {2018})}\BibitemShut {NoStop}%
	\bibitem [{\citenamefont {Arrasmith}\ \emph {et~al.}(2021)\citenamefont
		{Arrasmith}, \citenamefont {Cerezo}, \citenamefont {Czarnik}, \citenamefont
		{Cincio},\ and\ \citenamefont {Coles}}]{arrasmith2021effect}%
	\BibitemOpen
	\bibfield  {author} {\bibinfo {author} {\bibfnamefont {A.}~\bibnamefont
			{Arrasmith}}, \bibinfo {author} {\bibfnamefont {M.}~\bibnamefont {Cerezo}},
		\bibinfo {author} {\bibfnamefont {P.}~\bibnamefont {Czarnik}}, \bibinfo
		{author} {\bibfnamefont {L.}~\bibnamefont {Cincio}},\ and\ \bibinfo {author}
		{\bibfnamefont {P.~J.}\ \bibnamefont {Coles}},\ }\bibfield  {title} {\bibinfo
		{title} {Effect of barren plateaus on gradient-free optimization},\ }\href
	{https://quantum-journal.org/papers/q-2021-10-05-558/} {\bibfield  {journal}
		{\bibinfo  {journal} {Quantum}\ }\textbf {\bibinfo {volume} {5}},\ \bibinfo
		{pages} {558} (\bibinfo {year} {2021})}\BibitemShut {NoStop}%
	\bibitem [{\citenamefont {Ruder}(2016)}]{ruder2016overview}%
	\BibitemOpen
	\bibfield  {author} {\bibinfo {author} {\bibfnamefont {S.}~\bibnamefont
			{Ruder}},\ }\bibfield  {title} {\bibinfo {title} {An overview of gradient
			descent optimization algorithms},\ }\href@noop {} {\bibfield  {journal}
		{\bibinfo  {journal} {preprint arXiv:1609.04747}\ } (\bibinfo {year}
		{2016})}\BibitemShut {NoStop}%
	\bibitem [{\citenamefont {Powell}(1970)}]{powell1970survey}%
	\BibitemOpen
	\bibfield  {author} {\bibinfo {author} {\bibfnamefont {M.}~\bibnamefont
			{Powell}},\ }\bibfield  {title} {\bibinfo {title} {A survey of numerical
			methods for unconstrained optimization},\ }\href
	{https://epubs.siam.org/doi/abs/10.1137/1012004} {\bibfield  {journal}
		{\bibinfo  {journal} {SIAM review}\ }\textbf {\bibinfo {volume} {12}},\
		\bibinfo {pages} {79} (\bibinfo {year} {1970})}\BibitemShut {NoStop}%
	\bibitem [{\citenamefont {Grefenstette}(1993)}]{grefenstette1993genetic}%
	\BibitemOpen
	\bibfield  {author} {\bibinfo {author} {\bibfnamefont {J.~J.}\ \bibnamefont
			{Grefenstette}},\ }\bibfield  {title} {\bibinfo {title} {Genetic algorithms
			and machine learning},\ }in\ \href
	{https://dl.acm.org/doi/pdf/10.1145/168304.168305} {\emph {\bibinfo
			{booktitle} {Proceedings of the sixth annual conference on Computational
				learning theory}}}\ (\bibinfo {year} {1993})\ pp.\ \bibinfo {pages}
	{3--4}\BibitemShut {NoStop}%
	\bibitem [{\citenamefont {Giovannetti}\ \emph {et~al.}(2011)\citenamefont
		{Giovannetti}, \citenamefont {Lloyd},\ and\ \citenamefont
		{Maccone}}]{giovannetti2011advances}%
	\BibitemOpen
	\bibfield  {author} {\bibinfo {author} {\bibfnamefont {V.}~\bibnamefont
			{Giovannetti}}, \bibinfo {author} {\bibfnamefont {S.}~\bibnamefont {Lloyd}},\
		and\ \bibinfo {author} {\bibfnamefont {L.}~\bibnamefont {Maccone}},\
	}\bibfield  {title} {\bibinfo {title} {Advances in quantum metrology},\
	}\href {https://www.nature.com/articles/nphoton.2011.35} {\bibfield
		{journal} {\bibinfo  {journal} {Nat. Photonics}\ }\textbf {\bibinfo {volume}
			{5}},\ \bibinfo {pages} {222} (\bibinfo {year} {2011})}\BibitemShut {NoStop}%
	\bibitem [{\citenamefont {Jamio{\l}kowski}(1972)}]{jamiolkowski1972linear}%
	\BibitemOpen
	\bibfield  {author} {\bibinfo {author} {\bibfnamefont {A.}~\bibnamefont
			{Jamio{\l}kowski}},\ }\bibfield  {title} {\bibinfo {title} {Linear
			transformations which preserve trace and positive semidefiniteness of
			operators},\ }\href
	{https://www.sciencedirect.com/science/article/abs/pii/0034487772900110}
	{\bibfield  {journal} {\bibinfo  {journal} {Reports on mathematical physics}\
		}\textbf {\bibinfo {volume} {3}},\ \bibinfo {pages} {275} (\bibinfo {year}
		{1972})}\BibitemShut {NoStop}%
	\bibitem [{\citenamefont {Choi}(1975)}]{choi1975completely}%
	\BibitemOpen
	\bibfield  {author} {\bibinfo {author} {\bibfnamefont {M.-D.}\ \bibnamefont
			{Choi}},\ }\bibfield  {title} {\bibinfo {title} {Completely positive linear
			maps on complex matrices},\ }\href {Linear transformations which preserve
		trace and positive semidefiniteness of operators} {\bibfield  {journal}
		{\bibinfo  {journal} {Linear algebra and its applications}\ }\textbf
		{\bibinfo {volume} {10}},\ \bibinfo {pages} {285} (\bibinfo {year}
		{1975})}\BibitemShut {NoStop}%
	\bibitem [{zen()}]{zenodo}%
	\BibitemOpen
	\href {https://zenodo.org/records/16207268} {\bibinfo {title}
		{https://zenodo.org/records/16207268}}\BibitemShut {NoStop}%
\end{thebibliography}
\end{document}